\documentclass[journal,twoside]{IEEEtran}
\pdfoutput=1
\usepackage[english]{babel}
\usepackage{graphicx}
\usepackage{subfigure}
\usepackage{multicol}
\usepackage{setspace}
\usepackage{amsmath}
\usepackage{epstopdf}
\usepackage{fancyhdr}
\usepackage{indentfirst}
\usepackage{enumerate}
\usepackage{caption}
\usepackage{leftidx}
\usepackage{amssymb}
\usepackage{float}
\usepackage{breqn}
\usepackage{bm}
\usepackage{bbm}
\usepackage{booktabs}

\usepackage{mathrsfs}
\usepackage{cite}
\usepackage{multirow}
\usepackage{totcount}
\usepackage{algorithm}
\usepackage{algorithmic}
\usepackage{color}
\usepackage{makecell}
\usepackage{textcomp}
\allowdisplaybreaks[4]
\begin{document}

%\title{Model-Driven Deep Networks for Random Access in LEO Satellite Communications}

\title{Random Access with Massive MIMO-OTFS \\ in LEO Satellite Communications}
\author{Boxiao Shen, ~\IEEEmembership{Student Member, ~IEEE}, Yongpeng Wu, ~\IEEEmembership{Senior Member, ~IEEE}, Jianping An, ~\IEEEmembership{Member, ~IEEE}, Chengwen Xing, ~\IEEEmembership{Member, ~IEEE}, Lian Zhao, ~\IEEEmembership{Senior Member, ~IEEE}, and Wenjun Zhang, ~\IEEEmembership{Fellow, ~IEEE}
	%\thanks{The work was supported in part} 
	\thanks{This work was supported in part by the National Key R\&D Program of China under Grant 2018YFB1801102, National Science Foundation (NSFC) under Grant 62122052 and 62071289, and National Natural Science Foundation of China under Grant 62071398. (Corresponding authors: Yongpeng Wu and Jianping An.)}
	\thanks{B. Shen, Y. Wu, and W. Zhang are with the Department of Electronic Engineering, Shanghai Jiao Tong
		University, Shanghai 200240, China (e-mail:
		boxiao.shen@sjtu.edu.cn; yongpeng.wu@sjtu.edu.cn; zhangwenjun@sjtu.edu.cn).
		}
	\thanks{J. An and C. Xing are with the School of Information and
		Electronics, Beijing Institute of Technology, Beijing 100081, China (e-mails: an@bit.edu.cn;  xingchengwen@gmail.com).}
	\thanks{L. Zhao is with the Department of Electrical, Computer, and Biomedical Engineering,
		Ryerson University, Toronto, ON M5B 2K3, Canada (e-mail: l5zhao@ryerson.ca).}
}

\maketitle

\IEEEpeerreviewmaketitle
\begin{abstract}
This paper considers the joint channel estimation and device activity detection in the grant-free random access systems, where a large number of Internet-of-Things devices intend to communicate with a low-earth orbit satellite in a sporadic way. In addition, the massive multiple-input multiple-output (MIMO) with orthogonal time-frequency space (OTFS) modulation is adopted to combat the dynamics of the terrestrial-satellite link. We first analyze the input-output relationship of the single-input single-output OTFS when the large delay and Doppler shift both exist, and then extend it to the grant-free random access with massive MIMO-OTFS. Next, by exploring the sparsity of channel in the delay-Doppler-angle domain, a two-dimensional pattern coupled hierarchical prior with the sparse Bayesian learning and covariance-free method (TDSBL-CF) is developed for the channel estimation. Then, the active devices are detected by computing the energy of the estimated channel. Finally, the generalized approximate message passing algorithm combined with the sparse Bayesian learning and two-dimensional convolution (ConvSBL-GAMP) is proposed to decrease the computations of the TDSBL-CF algorithm. Simulation results demonstrate that the proposed algorithms outperform conventional methods.

\end{abstract}

\section{Introduction}
\IEEEPARstart{S}{atellite} communication technology has been widely concerned since its inception, which will help to expand terrestrial communication networks due to its characteristics of large coverage and robustness for different geographical conditions. A lot of standardization works on satellite communications are currently underway\cite{s39}, some of which hope to incorporate it into 5G and future networks\cite{s52}. Among the application requirements of 5G, the massive machine-type communications (mMTC) is an important scenario and a basic part of Internet-of-Things (IoT) applications\cite{s40}. A considerable part of the IoT devices is distributed in remote areas, such as deserts, oceans, forests, etc\cite{s41}. However, the existing cellular communication networks are mainly deployed in places where populations are concentrated, and most of them are for communications between people. As a result, supporting remote mMTC is still a huge challenge for the existing communication facilities. Fortunately, satellite communications provide a very promising solution, especially those based on low-earth orbit (LEO) satellites. The devices can access it with a low propagation delay, low path loss, and flexible elevation angle, compared with the traditional geostationary and medium earth orbit satellite \cite{ll87}. Until now, there have been many companies planning to launch LEO satellites to provide seamless network coverage to the ground users, such as SpaceX\cite{spacex} and OneWeb\cite{oneweb}. In addition, the grant-free random access (RA) is a very natural solution for satellite IoT communications\cite{s411,s53} when considering the features of the mMTC, i.e, the number of potential devices is huge while only a small part of the devices are active in any period of time and the amount of data sent by each device is small. Therefore, investigating grant-free RA technology is significant for satellite-enabled IoT systems. 

Over the past few years, bunches of methods have been proposed for channel estimation and device activity detection in the grant-free RA systems, aiming at Rayleigh fading channel \cite{s31,s32,s33,s34,s36,s37}, Rician fading channel\cite{s38} and Gaussian mixture model \cite{dl1}. Specifically, in \cite{s31,s32,s33,s34}, the authors transformed the joint channel estimation and device activity detection into a sparse signal recovery problem in compressed sensing, where the pilots of all users form a sensing matrix, and then the approximate message passing (AMP) algorithm\cite{s35} is used to solve it. The authors in \cite{s36,s37} adopted the variational Bayesian inference to solve this problem. In\cite{s38}, the authors considered both the line-of-sight (LoS) and the scattering component, and then proposed the Bernoulli–Rician message passing with expectation–maximization (EM) algorithm to detect active users and estimate their channels. In \cite{dl1}, to perform joint channel estimation and device activity detection, the authors adopted the Bernoulli-Gaussian mixture distribution as the prior of channel, and designed a neural network architecture by unrolling the vector AMP and combining the EM algorithm. The above literature only considered sparsity due to the sporadic device activity. When the massive multiple-input multiple-output (MIMO) is adopted in the systems, the channel will also be sparse in the angular domain. In \cite{s43}, the generalized multiple measurement vector AMP (GMMV-AMP) was proposed to explore the sparsity both in the device domain and angular domain, where the sparsity ratio of each coefficient in the channel matrix not only depends on itself but also on its neighborhoods along the angular axis. As a result, the performance of the estimation and detection is improved compared with the conventional methods. 
The above works are based on the block fading channel which is constant during one transmission. However, the high mobility of LEO satellites inevitably leads to the rapid change of terrestrial-satellite link (TSL), and the large Doppler shift will also degrade the communication quality. Therefore, the current grant-free random access schemes can not be directly applied for the LEO satellite communications.

Orthogonal time-frequency space (OTFS) is a promising solution
to tackle the time-variant channels\cite{s2,s4,s8,s21}, which converts the time-variant channels into the time-independent
channels in the delay-Doppler domain. There has been a lot of literature focusing on the applications of OTFS for point-to-point communications \cite{s9,s6,s18}, downlink transmissions \cite{s22,s20,s1}, and uplink transmissions \cite{s7,s23}. Specifically, in \cite{s9} and \cite{s6}, a general
framework of OTFS based on ideal pulse-shaping waveforms
was introduced for single-input single-output (SISO) systems. The authors in \cite{s18} analyzed the input-output relationship of OTFS with the ideal and rectangular waveforms, and adopted the message passing algorithm to detect the signal. In the applications of OTFS for the downlink transmissions, the authors in \cite{s22} derived the input-output relationship of OTFS with massive MIMO, and identified the 3D-structured sparsity of channel in the delay-Doppler-angle (DDA) domain. Then, considering the above sparsity, the 3D-structured orthogonal
matching pursuit algorithm was proposed to estimate the channel. The authors in \cite{s20} considered the existence of the fractional Doppler, and derived a more accurate channel model for the MIMO-OTFS. At the same time, the deterministic pilot design method was proposed. An uplink-aided high mobility downlink channel
estimation scheme for the massive MIMO-OTFS networks was proposed in \cite{s1}, where the EM-based variational Bayesian framework was adopted to estimate the parameters of the uplink channel, and then the angles, the delays, and the
Doppler frequencies were constructed for the downlink channels by exploring the reciprocity between the uplink and
the downlink channel. In the applications of OTFS for the uplink transmission, the authors in \cite{s7} proposed the non-orthogonal multiple-access transmission protocol that incorporated OTFS modulation for the heterogeneous user mobility. A new path division multiple access for both uplink and downlink massive MIMO-OTFS was proposed in \cite{s23}. Besides, as a multicarrier modulation technology, OTFS has a lower peak-to-average power ratio (PAPR) compared with the orthogonal frequency division multiplexing (OFDM) when a small or moderate number of Doppler bins are adopted\cite{s4,m18}. This typically is the case in the LEO satellite communications that the number of Doppler bins should be small enough to guarantee the effectiveness of the estimated channel due to the fast variations of the TSL. When a large number of Doppler bins are assumed, the PAPR of the OTFS is similar to that of the OFDM\cite{m18}, and then the methods proposed in \cite{m19,m20,m21} can be adopted to reduce the PAPR of OTFS.

In recent years, the MIMO transmission has been widely applied in the terrestrial wireless networks\cite{s45}, and massive MIMO can bring huge improvements in throughput and enhance spectral efficiency\cite{s46,s54}. However, due to the limited on-board processing capacity and instinct channel characteristics, employing the MIMO or even massive MIMO in satellite communications is still an open and challenging task\cite{s48}. Fortunately, there have been some efforts towards this goal. By exploiting the dual polarization, the two by two and four by two MIMO configurations were studied for the GEO satellite in \cite{m14}. The European Space Agency initiated the project to investigate the massive MIMO over satellite in \cite{esa}. A practical design of massive MIMO combined with the radio resource management
	approach was proposed in \cite{m15} to ease its implementation at the satellite payload level. The applications of massive MIMO along with full frequency reuse in the multi-user LEO satellite communications were considered in \cite{ss32,m13,m16,s47}. Motivated by this, we exploit the massive MIMO in the LEO satellite communications.

Overall, most of the previous grant-free random access schemes only consider the block fading channel and assume that the channel is constant during the pilot and the subsequent data transmission. However, time variation, multi-path propagation, and Doppler effects should also be involved in the TSL due to the high mobility of the LEO satellite\cite{ss32,mm2}, which results in our considered doubly dispersive channel\cite{s11}. Furthermore, there is little literature focusing on the grant-free random access with OTFS. In this paper, we investigate the joint channel estimation and device activity detection in the LEO satellite communications with massive MIMO-OTFS, where the OTFS is adopted to suppress the doubly dispersive effect in the TSL link and the massive MIMO is integrated for improving the performance. 
	Notice that the adoption of massive MIMO-OTFS will bring the 3D-structured sparsity in the delay-Doppler-angle domain\cite{s22}, which leads to the two-dimensional (2D) burst block sparsity of the channel matrix in the grant-free random access scenarios. This new type of sparsity is not studied in the previous literature and cannot be fully addressed by conventional algorithms. Therefore, we propose to deploy the sparse Bayesian learning (SBL) framework to capture the 2D burst block sparsity. Besides, the scale of our considered problem is much larger than that in the conventional literature, due to the estimation of the doubly dispersive channel. Thus, the algorithm based on the generalized approximate message passing (GAMP) is proposed to further decrease the computations of the SBL method.
%{\color{blue}In this paper, we investigate the joint channel estimation and device activity detection in the LEO communications with MIMO-OTFS, where the sparse Bayesian learning (SBL) framework is developed to capture the two-dimensional burst block sparsity, and then the algorithm based on the generalized approximate message passing (GAMP) is proposed to further decrease the computations of the SBL method.} 
%and exploring the channel sparsity in the delay-Doppler-angle domain with the uplink transmissions. In this paper, we adopt the massive MIMO-OTFS to the grant-free random access in LEO satellite communications to
%suppress the negative effect caused by the high-mobility of the satellite, and propose the joint channel estimation and device activity detection algorithms which utilize the channel sparsity in the delay-Doppler-angle domain. 
Our main contributions are summarized as follows:
\begin{itemize}
	\item We first analyze the input-output relationship of SISO-OTFS when the large delay and Doppler both exist. Then, we extend it to the grant-free random access system with massive MIMO-OTFS. We formulate the joint channel estimation and device activity detection as a sparse signal recovery problem, where the 3D-structured sparsity of the channel in the delay-Doppler-angle domain is transformed to the 2D burst block sparsity in our channel matrix.
	\item We propose a 2D pattern coupled hierarchical prior with the SBL framework to capture the 2D burst block sparsity of the channel matrix, where the sparsity pattern of each coefficient is controlled not only
	by its own hyperparameter, but also by the hyperparameters of its neighborhoods both along the delay-Doppler axis and angular axis in the channel matrix. Then, those hyperparameters and the precision of the noise are updated by the EM algorithm, and the covariance-free method is introduced to decrease the computational complexity of the SBL, resulting in the TDSBL-CF algorithm.
	\item To facilitate the TDSBL-CF algorithm for solving the large-scale problems, the GAMP with the 2D convolution and SBL (ConvSBL-GMAP) is developed. Here, we consider a special case of the proposed prior in the SBL framework to capture the 2D burst block sparsity. Then, based on this prior, the GAMP is utilized to estimate the posterior mean and variance of the channel matrix. At the same time, the update of the hyperparameters in the EM algorithm can be computed through the 2D convolution.
\end{itemize}

The rest of this paper is organized as follows. Section \ref{sp} investigates the system model for the massive MIMO-OTFS random access and formulates the problem. Section \ref{2} proposes the 2D pattern coupled SBL for joint channel estimation and device activity detection. Section \ref{3} combines the GAMP into the SBL framework to further decrease its complexity. Section \ref{4} evaluates the performance of the proposed algorithms.

\emph{Notations}: The superscripts $(\cdot)^{*}$ and $(\cdot)^{\mathrm{H}}$ denote the conjugate and conjugated-transpose operations, respectively. The boldface letters denote matrices or vectors. $\bar{\jmath}=\sqrt{-1}$ denotes the imaginary unit. $\|\mathbf{X}\|_{1}$ and $\|\mathbf{x}\|_{1}$ denote the $\ell_{1}$-norm of $\mathbf{X}$ and $\mathbf{x}$, $\|\mathbf{X}\|_{\text F}$ denotes the Frobenius norm of $\mathbf{X}$, and $\|\mathbf{x}\|_{2}$ denotes the $\ell_{2}$-norm of $\mathbf{x}$. $\lceil x\rceil$ denotes the smallest integer that is not less than $x$, while $\lfloor x\rfloor$ denotes the largest integer that is not greater than $x$. $\odot$ is the Hadamard product operator and $\oslash$ denotes element-wise division. $(\cdot)_{M}$ denotes mod $M$, and $\langle x\rangle_{N}$ denotes $\left(x+\left\lfloor\frac{N}{2}\right\rfloor\right)_{N}-\left\lfloor\frac{N}{2}\right\rfloor .$ The notation $\triangleq$ is used for definitions. $\delta(\cdot)$ denotes the Dirac delta function. $\mathbf{X}[a,:]$ denotes the $a$-th row of $\mathbf{X}$, wihle $\mathbf X[:,b]$ denotes the $b$-th column of $\mathbf{X}$. $X[a,b]$ and $X_{a,b}$ denote the $(a, b)$-th element of $\mathbf{X}$. $x_a$ denotes the $a$-th element of $\mathbf{x}$.

%\clearpage
\section{System Model}
\label{sp}
In this section, some basic concepts of OTFS are first reviewed. Then, we analyze the input-output relationship of the SISO-OTFS system, when both the large delay and Doppler shift exist. Next, an extension of OTFS into massive MIMO systems is described. Finally, we formulate the considered problem.
\subsection{SISO-OTFS Transmissions}
\subsubsection{Terrestrial Transmitter} 
The device in land employs the OTFS transmitter, which first maps a set of $NM$ information symbols $X^{\text{DD}}[k,l]$ in the delay-Doppler domain to the symbols $X^{\text{TF}}[n,m]$ in the time-frequnecy domain through the inverse symplectic finite Fourier transform (ISFFT)\cite{s21}, i.e., 
\begin{align}
\label{xx}
	X^{\text{TF}}[n, m]=\frac{1}{\sqrt{M N}} \sum_{k=\lceil-N / 2\rceil}^{\lceil N / 2\rceil-1} \sum_{l=0}^{M-1} X^{\mathrm{DD}}[k, l] e^{-\bar{\jmath} 2 \pi\left(\frac{m l}{M}-\frac{n k}{N}\right)},
\end{align}
where $k=\lceil -N/2\rceil,\dots,\lceil N/2\rceil-1$, $l=0,\dots,M-1$. Next, the OFDM modulator converts the symbols $X^{\text{TF}}[n,m]$ to a continuous signal $s(t)$ using a transmit waveform $g_{\text{tx}}(t)$ as
\begin{align}
	%\begin{array}{r}
	s(t)=\sum\limits_{m=0}^{M-1} \sum\limits_{n=0}^{N-1} X^{\mathrm{TF}}[n, m] e^{\bar{\jmath} 2 \pi m \Delta f\left(t-T_{\text{cp}}-n T_{\mathrm{sym}}\right)} \nonumber \\
	 \times g_{\mathrm{tx}}\left(t-n T_{\mathrm{sym}}\right),
	%\end{array}
\end{align}
where $\Delta f$ is the carrier spacing,  $T_{\text{cp}}=\frac{M_{\text{cp}} T}{M}$ is the time duration of the cyclic prefix (CP), $M_{\text{cp}}$ is the length of CP, $T=\frac{M T_{\mathrm{sym}}}{M+M_{\text{cp}}}$ is the time duration of an OFDM symbol without CP,  $T_{\mathrm{sym}}$ is the time duration of a complete OFDM symbol, and $g_{\text{tx}}(t)$ is defined as
\begin{align}
	g_{\mathrm{tx}}(t) \triangleq \begin{cases}\frac{1}{\sqrt{T}}, & 0 \leq t \leq T_{\mathrm{sym}} \\ 0, & \text { otherwise }\end{cases}.
\end{align}

\subsubsection{Terrestrial-satellite Link}
The TSLs will experience fast variations due to the high mobility of the LEO satellite, and the signal will be transmitted over LoS path and non-LoS (NLoS) path due to the
direct signal interacting with the scatterers in the vicinity
of the terrestrial device.  Therefore, the time variation,
multipath propagation and Doppler effects should be considered in the LEO satellite channel. Then, the channel response in the time-frequency domain can be represented as\cite{ss32,m13,m16,s47}
\begin{align}
\label{cm}
	h(t,f)=&\sqrt{\frac{\mathcal{K}}{\mathcal{K} + 1}}e^{\bar{\jmath} 2 \pi\left(t \nu^{\text{LoS}}-f \tau^{\text{LoS}}\right)} \nonumber \\
	&+ \sqrt{\frac{1}{\mathcal{K} + 1}}\sum_{i=1}^{P} h_i^{\text{NLoS}} e^{\bar{\jmath} 2 \pi\left(t \nu_{i}^{\text{NLoS}}-f \tau_{i}^{\text{NLoS}}\right)},
\end{align}
where $\mathcal{K}$ is the Rician factor; $\nu^{\text{LoS}}$ and $\tau^{\text{LoS}}$ are the Doppler shift and propagation delay of the LoS path, respectively; $h_i^{\text{NLoS}}$, $\nu_{i}^{\text{NLoS}}$, and $\tau_{i}^{\text{NLoS}}$ denote  the complex gain, Doppler shift, and propagation delay of the  $i$-th NLoS path, respectively. Note that the channel model in (\ref{cm}) is applicable when the relative positions of the LEO
satellite and device do not change significantly, and thus the physical parameters $\nu^{\text{LoS}}$, $\tau^{\text{LoS}}$, $h_i^{\text{NLoS}}$, $\nu_{i}^{\text{NLoS}}$, and $\tau_{i}^{\text{NLoS}}$ are assumed to be invariant\cite{ss32}. When the LEO satellite or the device moves over a long distance, those parameters should be updated accordingly\cite{wc}. Next, we elaborate the key parameters as follows.
\begin{itemize}
	\item \textbf{Doppler shift}: In the LEO satellite communications, the Doppler shift is much larger than that in the terrestrial wireless networks due to the large relative velocity between the satellite and the devices. For example, the Doppler shift can be up to 41 kHz at the 2 GHz carrier frequency for a LEO satellite operated at the 600 km altitude\cite{ss5}. Generally, the Doppler shift $\nu^{\text{LoS}}$ is composed of two parts, i.e., $\nu^{\text{LoS}}=\nu^{\text{LoS}}_{S} + \nu^{\text{LoS}}_{D}$, where $\nu^{\text{LoS}}_{S}$ and $\nu^{\text{LoS}}_{D}$ are the Doppler shifts concerned with the movement of the satellite and the device, respectively. For the NLoS path, similarly, we have $\nu_{i}^{\text{NLoS}}=\nu^{\text{NLoS}}_{S,i} + \nu^{\text{NLoS}}_{D,i}$. As a result of the high altitude of the satellite, the first part $\nu^{\text{LoS}}_{S}$ ($\nu_{S,i}^{\text{NLoS}}$) is nearly identical for different paths of the TSL\cite{s50}, and thus $\nu^{\text{LoS}}_{S}$ ($\nu_{S,i}^{\text{NLoS}}$) can be re-written as $\nu^{\text{LoS}}_{S} = \nu_{S,i}^{\text{NLoS}} = \nu_S$. On the other hand, the Doppler shift $\nu^{\text{LoS}}_D$ ($\nu^{\text{NLoS}}_{D,i}$) due to the motion
	of the device is typically distinct for different propagation	paths\cite{ss32,m13}.
	\item \textbf{Propagation delay}: The propagation delay also exhibits much larger value than that in the terrestrial wireless networks, because of the long distance between the satellite and the device. For example, the roundtrip delay is about 17.7 ms for an LEO satellite at the 1000 km altitude and with 45° of elevation angles\cite{s51}. 
\end{itemize}
To convert the time-varying channel into the time-independent channel, the Fourier transform and inverse Fourier transform are applied for the channel response $h(t,f)$ along the time dimension and frequency dimension, respectively, and then we have
\begin{align}
h(\tau,\nu)&= \iint h(t,f) e^{-\jmath2\pi(\nu t - f\tau )} dtdf\nonumber \\
&=\sqrt{\frac{\mathcal{K}}{\mathcal{K} + 1}}\delta\left(\tau-\tau^{\text{LoS}}\right) \delta\left(\nu-\nu^{\text{LoS}}\right) + \nonumber \\ 
&\sqrt{\frac{1}{\mathcal{K} + 1}}
\sum_{i=1}^{P}  h_{i}^{\text{NLoS}} \delta\left(\tau-\tau_{i}^{\text{NLoS}}\right) \delta\left(\nu-\nu_{i}^{\text{NLoS}}\right),
\end{align}
which is the channel response in the delay-Doppler domain. We denote $\sqrt{\frac{\mathcal{K}}{\mathcal{K} + 1}}$, $\nu^{\text{LoS}}$, $\tau^{\text{LoS}}$,  $\sqrt{\frac{1}{\mathcal{K} + 1}}  h_{i}^{\text{NLoS}}$, $\nu_{i}^{\text{NLoS}}$, and $\tau_{i}^{\text{NLoS}}$ as $h_0$, $\nu_0$, $\tau_0$, $h_{i}$, $\nu_{i}$, and $\tau_i$ for $i = 1\dots P$, respectively, and then the channel can be re-written in a more compact way, i.e., 
\begin{align}
\label{cc}
h(\tau, \nu)=\sum_{i=0}^{P} h_{i} \delta\left(\tau-\tau_{i}\right) \delta\left(\nu-\nu_{i}\right).
\end{align} 

%Since generally there are only a small number of reflectors around the IoT devices in the TSL, the channel in the delay-Doppler domain can be modelled by very few parameters\cite{s18}, and it’s reasonable to
%assume that there exists both of the LoS and a few
%non-LoS (NLoS) paths\cite{ss32}. Hence, the sparse representation of the channel $h(\tau, \nu)$ is given as
%where the first term represents the LoS path associated with the delay $\tau_{0}$ and Doppler shift $\nu_{0}$, and there are other $P$ terms corresponding to the NLoS paths associated with the path gain $\tilde h_{i}$, delay $\tau_{i}$ and Doppler shift $\nu_{i}$; , and $\delta(\cdot)$ is the Dirac delta function. 
Finally, we define the delay and Doppler taps for the $i$-th path as follows
\begin{align}
\label{tap}
	\tau_i=\frac{l_i + c_iM}{M\Delta f}, \nu_i = \frac{k_i + \tilde{k}_i + b_iN}{NT_{\text{sym}}},
\end{align}
where $l_i = 0,\dots,M-1$ and $k_i = \lceil -N/2\rceil,\dots,\lceil N/2 \rceil -1$ represent the indexes of the delay tap and Doppler tap corresponding to the delay $\tau_i$ and Doppler $\nu_i$, respectively; $b_i$ and $c_i$ are integers and referred to as the outer Doppler and outer delay, respectively, since they represent the part exceeding one carrier spacing and one symbol duration; $\tilde{k}_i \in (-\frac{1}{2},\frac{1}{2}]$ is the fractional Doppler shift. In addition, we define the tap index set $\mathcal{D}$ and $\mathcal{F}$ of the delay and Doppler shift of all paths as
$\mathcal{D} = [l_0,\dots,l_P]$, and  $\mathcal{F} = [k_0+\tilde{k}_0,\dots,k_P+\tilde{k}_P]$. 

 %The reason we define the outer cofficients is that the long delay and large Doppler shift will both exist in LEO communications[], and usually we can not choose suitable carrier spacing and symbol duration to suppress the two negative effects simultaneously.
\subsubsection{LEO Satellite Receiver}
The signal $s(t)$ is transmitted over the above TSL, and then the received signal $r(t)$ is given by\cite{s18}
\begin{align}
r(t)=\iint h(\tau, \nu) s(t-\tau) e^{j 2 \pi \nu(t-\tau)} d \tau d \nu + n(t),
\end{align}
where $n(t)$ is the additive Gaussian noise.

At the receiver, the cross-ambiguity function $A_{g_{\mathrm{rx}}, r}(t, f)$ is computed by the matched filter as follows
\begin{align}
	A_{g_{\mathrm{rx}}, r}(t, f) = \int g_{\mathrm{rx}}^{*}\left(t^{\prime}-t\right) r\left(t^{\prime}\right) e^{-\bar{\jmath} 2 \pi f\left(t^{\prime}-T_{\text{cp}}-t\right)} d t^{\prime},
\end{align}
where the received waveform $g_{\mathrm{rx}}(t)$ is defined as
\begin{align}
	g_{\mathrm{rx}}(t) \triangleq \begin{cases}\frac{1}{\sqrt{T}}, & T_{\text{cp}} \leq t \leq T_{\mathrm{sym}} \\ 0, & \text { otherwise }\end{cases}.
\end{align}
Then, the received symbols $Y^{\mathrm{TF}}[n, m]$ in the time-frequency domain are sampled from $A_{g_{\mathrm{rx}},r}(t, f)$ as
\begin{align}
\label{yy}
	Y^{\mathrm{TF}}[n, m]=\left.A_{g_{\mathrm{rx}}, r}(t, f)\right|_{t=n T_{\mathrm{sym}}, f=m \Delta f}.
\end{align}
Next, the symplectic finite Fourier transform (SFFT) is adopted to map $Y^{\mathrm{TF}}[n, m]$ to the symbols $Y^{\mathrm{DD}}[k, l]$ as 
\begin{align}
\label{yys}
	Y^{\mathrm{DD}}[k, l]=\frac{1}{\sqrt{N M}} \sum_{n=0}^{N-1} \sum_{m=0}^{M-1} Y^{\mathrm{TF}}[n, m] e^{\bar{\jmath} 2 \pi\left(\frac{m l}{M}-\frac{n k}{N}\right)}.
\end{align}
Similar to \cite{s20}, we define the complex gain of the time-variant channel on the delay tap $l$ at the time $\rho T_{\mathrm{s}}$ as
\begin{align}
	h_{\rho, l}=\sum_{i=0}^{P} h_{{i}} e^{\bar{\jmath} 2 \pi(\rho-l) T_{s} \nu_i} \delta\left(l T_{\mathrm{s}}-(\tau_{i})_T\right),
\end{align}
where $T_{\mathrm{s}}=\frac{1}{M \Delta f}$ is the system sampling interval. Then the input-ouput relationship of the SISO-OTFS system is shown in the following proposition. 

\textbf{Proposition 1:} For a SISO-OTFS system, when the time duration of CP is beyond the maximum delay of all paths, i.e., $T_{\text{cp}} \geq \tau_i, i = 0,\dots,P$, the input-output
relationship in the delay-Doppler domain is given by
\begin{align}
\label{in_out}
	Y^{\mathrm{DD}}[k, l]= &\sum_{l^{\prime}=0}^{M-1} \sum_{k^{\prime}=\lceil-N / 2\rceil}^{\lceil N / 2\rceil-1} \phi(l,l^{\prime})H^{\mathrm{DD}}\left[k^{\prime}, l^{\prime}\right] \nonumber \\
	&\times X^{\mathrm{DD}}\left[\left\langle k-k^{\prime}\right\rangle_{N},\left(l-l^{\prime}\right)_{M}\right] + Z^{\mathrm{DD}}[k, l],
\end{align}
where 
\begin{align}
\label{hdef}
	&H^{\mathrm{DD}}[k, l]=\frac{1}{N} \sum_{j=0}^{N-1} h_{M_{\text{cp}}+j(M+M_{\text{cp}),l}}  e^{-\bar{\jmath} \frac{2 \pi}{N} k j}, \\
	&\phi\left(l, l^{\prime}\right)= \begin{cases}e^{\bar{\jmath} 2 \pi \frac{\left(k_{i}+\tilde{k}_{i} + b_iN\right)\left(l-c_iM\right)}{\left(M+M_{\mathrm{cp}}\right) N}}, & l^{\prime} \in \mathcal{D}, l^{\prime}=l_{i} \\ 1, & l^{\prime} \notin \mathcal{D}\end{cases},
\end{align}
$l_{i}$ is selected from $\mathcal{D}$, $k_{i}+\tilde{k}_{i}$ is in $\mathcal{F}$, and  $Z^{\mathrm{DD}}[k, l]$ is the Gaussian noise.

\emph{Proof}: See Appendix \ref{p1}.

We can find in (\ref{in_out}) that the outer Doppler $b_i$ and outer delay $c_i$ appear in the phase compensation term $\phi\left(l, l^{\prime}\right)$, which means that they will cause the phase rotation of the received symbols in the delay-Doppler domain and may degrade the detection performance. Besides, if $\tau_i \ll T$ and $\nu_i \ll \Delta f$, $b_i$ and $c_i$ can be neglected, and then (\ref{in_out}) is identical to that in \cite{s20}. Notice that in (\ref{in_out}), we require $T_{\text{cp}} \geq \tau_i$ to eliminate the intersymbol interference (ISI) between the OFDM symbols in one OTFS frame and to reduce the complexity of the receiver when the large differential delay exist\cite{ss5}. However, this will lead to the low spectral efficiency since the longer CP is usually needed. For example, in the narrowband-IoT reference scenarios
from 3GPP\cite{3gpp}, the satellite operates in the S-band (2 GHz) and at the altitude 600 km. Then, the maximum differential delay and Dopppler shift seen in the uplink can be up to 1397 \textmu s and 41 kHz, respectively. As a result, the length of CP will be several multiples of the data duration (e.g., the carrier spacing 3.75 kHz\cite{3gpp} is adopted) if we let $ T_{\text{cp}} >$ 1397 \textmu s. Besides, the estimated CSI may be outdated soon due to dynamics of the TSL and the long symbol duration. Therefore, to improve the spectral effciency further, we assumed that the pre-compensation for delay at the device level using the initial downlink time synchronization algorithms\cite{f1} is performed such that the residual delay seen in the uplink is in a small range. Then, we can adopt a relatively short CP, and select the proper symbol time duration and carrier spacing such that $\tau_i \ll T$ and $\nu_i \ll \Delta f$. In this scenario, the phase compensation term is given by
\begin{align}
\label{phi_s}
\phi\left(l, l^{\prime}\right)= \begin{cases}e^{\bar{\jmath} 2 \pi \frac{\left(k_{i}+\tilde{k}_{i}\right)l}{\left(M+M_{\mathrm{cp}}\right) N}}, & l^{\prime} \in \mathcal{D}, l^{\prime}=l_{i} \\ 1, & l^{\prime} \notin \mathcal{D}\end{cases},
\end{align}
In the follwing content, we focus on the case that $\tau_i \ll T$ and $\nu_i \ll \Delta f$, where $\tau_i$ represents the residual delay seen in the uplink. 

\subsection{Random Access with Massive MIMO-OTFS}
\begin{figure}[!htb]
	\centering
	\includegraphics[width=2.45in]{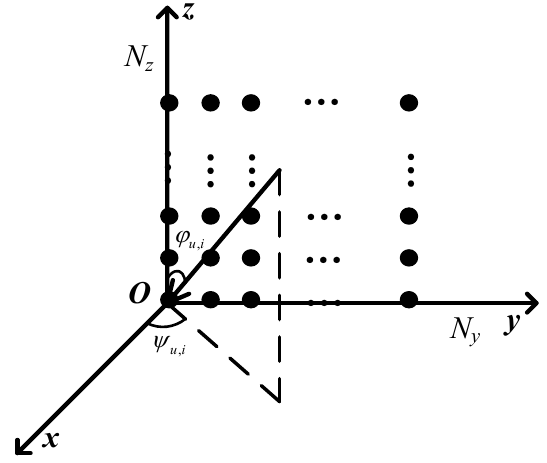}
	\caption{Illustration of UPA in the LEO satellite. Each black dot represents one antenna. }
	\label{antenna}
\end{figure} 

We consider the uplink transmission in a massive MIMO-OTFS system, where the LEO satellite is equipped with an uniform planar array (UPA) composed of $N_a = N_y \times N_z$ antennas, as illutrated in Fig. \ref{antenna}, where $N_y$ and $N_z$ are the number of antennas along the y-axis and z-axis, respectively, and the antenna spacing is set to half
wavelength. In a given time interval, there are $U$ potential IoT devices with single antenna transmitting the data to the satellite in a sporadic traffic way, i.e., only $U_a$ devices are active and $U_a \ll U$. In addition, the active devices share the same resources. We denote the activity indicator of the $u$-th device as $\lambda_u$, which is equal to 1 when it is active, and 0 otherwise. Then, the set of active IoT devices are represented as $\mathcal{U} = \{u|\lambda_u = 1, 0\leq u\leq U-1\}$, and the number of active IoT devices $U_a = \left|\mathcal{U}\right|$.
Similar to (\ref{in_out}) and (\ref{hdef}), at the satellite receiver, the symbols transmitted from the $u$-th device to the $(n_z + 1 + n_yN_z)$-th antenna in the delay-Doppler-space domain can be expressed as
\begin{align}
\label{in_out_singleU}
&Y_u^{\mathrm{DDS}}[k, l, n_y, n_z] \nonumber\\
&=\sum_{l^{\prime}=0}^{M-1} \sum_{k^{\prime}=\lceil-N / 2\rceil}^{\lceil N / 2\rceil-1} \phi_u(l,l^{\prime})H_u^{\mathrm{DDS}}\left[k^{\prime}, l^{\prime}, n_y,n_z\right] \nonumber\\
& \times X_u^{\mathrm{DD}}\left[\left\langle k-k^{\prime}\right\rangle_{N},\left(l-l^{\prime}\right)_{M}\right] + Z^{\mathrm{DD}}[k, l,n_y, n_z],
\end{align}
where $n_y = 0,\dots,N_y-1$, $n_z = 0,\dots,N_z-1$, $\phi_u(l,l^{\prime})$ is the phase compensation factor of the $u$-th device, $X_u^{\mathrm{DD}}\left[\left\langle k-k^{\prime}\right\rangle_{N},\left(l-l^{\prime}\right)_{M}\right]$ is the transmitted symbol of the $u$-th device in the delay-Doppler domain, and $Z^{\mathrm{DD}}[k, l,n_y, n_z]\sim \mathcal{CN}(0,\theta^{-1})$ is the noise with the precision $\theta$; $H_u^{\mathrm{DDS}}\left[k^{\prime}, l^{\prime}, n_y,n_z\right]$ is the delay-Doppler-space domain channel, which is defined as\cite{wc}
\begin{align}
\label{spaceh}
&H_u^{\mathrm{DDS}}[k, l,n_y,n_z]\nonumber\\
&=\frac{1}{N}\sum_{i=0}^{P} \sum_{j=0}^{N-1}  h_{u,i} e^{\bar{\jmath} 2 \pi\left(M_{\mathrm{cp}}+j\left(M+M_{\mathrm{cp}}\right)-l\right) T_{\mathrm{s}} \nu_{u,i}}\nonumber\\ 
&\times\delta\left(l T_{\mathrm{s}}-\tau_{u,i}\right) e^{\bar{\jmath} \pi n_y \sin \varphi_{u,i} \sin \psi_{u,i}} e^{\bar{\jmath} \pi n_z \cos \varphi_{u,i}} e^{-\bar{\jmath}  \frac{2 \pi}{N} k j},
\end{align}
where $h_{u,i}$, $\nu_{u,i}$, and $\tau_{u,i}$ are the complex channel gain, Doppler shift, and propagation delay of the $u$-th device in the $i$-th path, respectively; $\psi_{u,i}\in[-\pi/2,\pi/2]$ and $\varphi_{u,i}\in[0,\pi]$ are the corresponding azimuth angle and elevation angle, respectively. Since the satellite always operates on the relatively high altitude compared with that of the scatters located around the IoT devices, the angles
of all paths associated with the same device can be
assumed to be identical\cite{ss32}. Hence, we omit the path index of the azimuth and elevation angles in the following content.
 
To exploit the sparsity of the angular domain, the two-dimensional discrete Fourier transform (2D-DFT) is applyed for the channel of $u$-th device along the space dimension in the delay-Doppler-space domain, and then the delay-Doppler-angle domain channel is given by
\begin{align}
\label{angleD}
&H_u^{\mathrm{DDA}}[k, l, a_y,a_z] \nonumber\\
&=\frac{1}{\sqrt{N_yN_z}} \sum_{n_y=0}^{N_y-1}\sum_{n_z=0}^{N_z-1} H_u^{\mathrm{DDS}}\left[k, l, n_y,n_z\right] e^{-\bar{\jmath} 2 \pi (\frac{a_y n_y}{N_y}+\frac{a_z n_z}{N_z})} \nonumber\\
&=\sqrt{N_yN_z}\sum_{i=0}^{P}h_{u,i} e^{\bar{\jmath} 2 \pi(M_{\text{cp}} -l)T_s\nu_{u,i}} \Pi_N(k-NT_{\text{sym}}\nu_{u,i})\nonumber\\  &\times\Pi_{N_y}(a_y-N_y\Omega_{y_{u}}/2) \Pi_{N_z}(a_z-N_z\Omega_{z_{u}}/2)\delta\left(l T_{\mathrm{s}}-\tau_{u,i}\right),
\end{align}
where $a_y = 0,\dots,N_y-1$ and $a_z = 0,\dots,N_z-1$ are indexes along the angular domain; $\Omega_{y_{u,i}} = \sin \varphi_{u} \sin \psi_{u}$ and $\Omega_{z_{u}} = \cos \varphi_{u} $ are the directional cosines along the $y$-axis and $z$-axis, respectively; $\Pi_N(x)\triangleq\frac{1}{N} \sum_{i=0}^{N-1} e^{-\bar{\jmath} 2 \pi \frac{x}{N} i}$. Therefore, we can find in (\ref{angleD}) that $H_u^{\mathrm{DDA}}[k, l, a_y,a_z]$ has domiant elements only if $k \approx NT_{\text{sym}}\nu_{u,i}$, $l \approx \tau_{u,i} M \Delta f$, $a_y \approx N_y\Omega_{y_{u}}/2$, and $a_z \approx N_z\Omega_{z_{u}}/2$, which means that the channel in the delay-Doppler-angle domain has the 3D-structured sparsity\cite{s22}. In practice, the number of dominant paths $P\ll M$, and therefore, we will get very sparse channel in the delay-Doppler-angle domain. Combining (\ref{in_out_singleU}), (\ref{spaceh}), and (\ref{angleD}), the received symbols in the delay-Doppler-angle domain is given by
\begin{align}
&Y_u^{\mathrm{DDA}}[k, l, a_y, a_z] \nonumber\\
&= \frac{1}{\sqrt{N_yN_z}} \sum_{n_y=0}^{N_y-1}\sum_{n_z=0}^{N_z-1} Y_u^{\mathrm{DDS}}[k, l, n_y, n_z] e^{-\bar{\jmath} 2 \pi (\frac{a_y n_y}{N_y}+\frac{a_z n_z}{N_z})} \nonumber\\
&=\sum_{l^{\prime}=0}^{M-1} \sum_{k^{\prime}=\lceil-N / 2\rceil}^{\lceil N / 2\rceil-1} \phi_u(l,l^{\prime})H_u^{\mathrm{DDA}}\left[k^{\prime}, l^{\prime}, a_y,a_z\right] \nonumber\\
& \times X_u^{\mathrm{DD}}\left[\left\langle k-k^{\prime}\right\rangle_{N},\left(l-l^{\prime}\right)_{M}\right] + Z^{\mathrm{DDA}}[k, l,a_y, a_z],
\end{align}
where $Z^{\mathrm{DDA}}[k, l,a_y, a_z]$ is the Gaussian noise in the delay-Doppler-angle domain, obeying the same distribution as  $Z^{\mathrm{DDS}}[k, l,n_y, n_z]$.
Then, according to the superposition principle, the received symbols from all active devices can be expressed as 
\begin{align}
\label{allsig}
&Y^{\mathrm{DDA}}[k, l, a_y, a_z]\nonumber\\ &=\sum_{u=0}^{U-1}\sum_{l^{\prime}=0}^{M-1} \sum_{k^{\prime}=\lceil-N / 2\rceil}^{\lceil N / 2\rceil-1} \lambda_u\phi_u(l,l^{\prime})H_u^{\mathrm{DDA}}\left[k^{\prime}, l^{\prime}, a_y,a_z\right] \nonumber\\
&\times X_u^{\mathrm{DD}}\left[\left\langle k-k^{\prime}\right\rangle_{N},\left(l-l^{\prime}\right)_{M}\right] + Z^{\mathrm{DDA}}[k, l,a_y, a_z].
\end{align}  

\subsection{Problem Formulation}
%\begin{figure}[!htb]
	%\centering
	%\includegraphics[width=2in]{figure/system/pilot.pdf}
	%\caption{The position of pilots, data, and guard intervals in one OTFS frame.
	%}
	%\label{pilot}
%\end{figure} 
In this subsection, we formulate the joint device activity detection and channel estimation. 
%The pilot structure of an arbitrary device in the delay-Doppler domain is shown in Fig. \ref{pilot}. 
The length of pilots along the Doppler dimension and the delay dimension are $N_\nu$ and $M_\tau$, respectively. In addition, the guard intervals are inserted in one OTFS frame to avoid the interference between the data and pliots. Note that the channel in the delay-Doppler-angle domain has finite support $[0,M_{\text{max}}-1]$ along the delay dimension. The length of guard intervals should be $M_g\geq M_{\text{max}} - 1$, and we assume that $M_\tau\geq M_g$. Further, to fully explore the sparsity in the Doppler domain, we propose to utilize all the available resources in the Doppler domain to transmit pilots, i.e., set $N_\nu=N$. Finally, the non-orthogonal pilot pattern is adopted, and the training pilots are the complex Gaussian random sequences, which are independent among different devices. Then, in a given time interval, the active devices transmit their pilots simultaneously. We denote the pilots of $u$-th device as $X_{p,u}^{\mathrm{DD}}\left[\left\langle k-k^{\prime}\right\rangle_{N},\left(l-l^{\prime}\right)_{M}\right]$.
Then, according to (\ref{allsig}), the received pilots in the delay-Doppler-angle domain at the satellite side is given by
\begin{align}
\label{rpilot}
&Y_p^{\mathrm{DDA}}[k, l, a_y, a_z]\nonumber\\ &=\sum_{u=0}^{U-1}\sum_{l^{\prime}=0}^{M_\tau-1} \sum_{k^{\prime}=\lceil-N / 2\rceil}^{\lceil N / 2\rceil-1} \lambda_u\phi_u(l,l^{\prime})H_u^{\mathrm{DDA}}\left[k^{\prime}, l^{\prime}, a_y,a_z\right] \nonumber\\
&\times X_{p,u}^{\mathrm{DD}}\left[\left\langle k-k^{\prime}\right\rangle_{N},\left(l-l^{\prime}\right)_{M}\right] + Z^{\mathrm{DDA}}[k, l,a_y, a_z].
\end{align}
where $k=\lceil -N/2\rceil,\dots,\lceil N/2\rceil-1$, and $l=0,\dots,M_\tau-1$. 

To facilitate the following analysis, we rewrite (\ref{rpilot}) into the matrix form as
\begin{align}
\label{incompletey}
	\mathbf Y = (\Phi \odot \mathbf X)\Lambda \tilde{\mathbf H} + \mathbf Z,
\end{align}
where the elements of Z are independent Gaussian noise, and the $(a_z+1+N_za_y)$-th column of $\mathbf Y$ is $\mathrm{vec}\left(\mathbf Y_p^{DDA}[:,:,a_y, a_z]\right)$. $\Phi=[\Phi_0,\dots,\Phi_{U-1}]$ contains the phase compensation matrices $\Phi_u\in \mathrm{C}^{M_\tau N\times UM_\tau N}$ of the $u$-th device composed of $\phi_u(l,l)\mathbf{1}_N$ and $l=0,\dots,M_\tau-1$.
%\begin{align}
	%\Phi_u=\left[\begin{array}{cccc}
	%\phi_u[0,0]\mathbf{1}_N &  \cdots & %\phi_u[0,M_\tau-1]\mathbf{1}_N \\
	%\phi_u[1,0]\mathbf{1}_N &  \cdots & %\phi_u[1,M_\tau-1]\mathbf{1}_N \\
	%\vdots & \vdots &  \vdots \\
	%\phi_u[M_\tau-1,0]\mathbf{1}_N &  \cdots & %\phi_u[M_\tau-1,M_\tau-1]\mathbf{1}_N
	%\end{array}\right].
%\end{align}
%where $\mathbf{1}_N\in \mathrm{R}^{N\times N}$ with all elements that is equal to one.
$\mathbf X=[\mathbf X_0,\dots,\mathbf X_{U-1}]\in \mathrm{C}^{M_\tau N\times UM_\tau N}$, where $\mathbf X_u\in\mathrm{C}^{M_\tau N\times M_\tau N}$ is the pilot matrix of the $u$-th device. Note that $\mathbf X_u$ is a doubly circulant matrix due to the 2D convolution in (\ref{rpilot}), %given by
%\begin{align}
	%\mathbf X_u=\left[\begin{array}{cccc}
	%\mathbf X_{u, 0} & \mathbf X_{u, M_\tau-1} & \cdots & %\mathbf X_{u, 1} \\
	%\mathbf X_{u, 1} & \mathbf X_{u, 0} & \cdots & \mathbf X_{u, 2} \\
	%\vdots & \vdots & \vdots & \vdots \\
	%\mathbf X_{u, M_\tau-1} & \mathbf X_{u, M_\tau-2} & \cdots & \mathbf X_{u, 0}
	%\end{array}\right],
%\end{align}
and its sub-matrix $\mathbf X_{u,l}$, $l=0,\dots,M_\tau-1$ is the circulant matrix, formed by $\left[X_{p,u}^{\mathrm{DD}}[0,l],X_{p,u}^{\mathrm{DD}}[1,l],\dots,X_{p,u}^{\mathrm{DD}}[-1,l]\right]$. 
%\begin{align}
	%\mathbf X_{u,l}=\left[\begin{array}{cccc}
	%X_{p,u}^{\mathrm{DD}}[0,l] & X_{p,u}^{\mathrm{DD}}[-1,l] & \cdots & X_{p,u}^{\mathrm{DD}}[1,l] \\
	%X_{p,u}^{\mathrm{DD}}[1,l] & X_{p,u}^{\mathrm{DD}}[0,l] & \cdots & X_{p,u}^{\mathrm{DD}}[2,l] \\
	%\vdots & \vdots & \vdots & \vdots \\
	%X_{p,u}^{\mathrm{DD}}[-1,l] & X_{p,u}^{\mathrm{DD}}[-2,l] & \cdots & X_{p,u}^{\mathrm{DD}}[0,l]
	%\end{array}\right].
%\end{align}
$\Lambda$ is a diagonal matrix composed of the activity indicator of all the devices, given by
\begin{align}
	\Lambda=\left[\begin{array}{llll}
	\lambda_0\mathbf I & & & \\
	&\lambda_1\mathbf I & & \\
	& & \ddots & \\
	& & & \lambda_{U-1}\mathbf I
	\end{array}\right],
\end{align}
where $\mathbf I$ is the identity matrix with dimension $M_\tau N\times M_\tau N$. Finally, the $(a_z+1+N_za_y)$-th column of $\tilde{\mathbf H}$ is given by
%from all potential devices, and its $(a_z+1+N_za_y)$-th column is given by
\begin{align}
\label{H_vec}
	\tilde{\mathbf H}[:,a_z+1+N_za_y]=\left[\begin{array}{cccc}
	\mathrm{vec}\left(\mathbf H^{\mathrm{DDA}}_0[:,:,a_y,a_z]\right) \\
	\mathrm{vec}\left(\mathbf H^{\mathrm{DDA}}_1[:,:,a_y,a_z]\right) \\
	\vdots \\
	 \mathrm{vec}\left(\mathbf H^{\mathrm{DDA}}_{U-1}[:,:,a_y,a_z]\right)
	\end{array}\right].
\end{align}

We denote the channel matrix $\Lambda \tilde{\mathbf H}$ as $\mathbf H$. Then, (\ref{incompletey}) can be written as
\begin{align}
\label{matrixform}
	\mathbf Y =(\mathbf{\Phi}\odot \mathbf X)\mathbf H + \mathbf Z.
\end{align}
Therefore, the joint device activity detection and channel estimation in the massive MIMO-OTFS systems is a sparse signal reconstruction problem. However, since the phase compensation matrix $\mathbf{\Phi}$ is related to the channel, and $\mathbf H$ cannot be directly recovered with the unknown sensing matrix. Fortunately, when choosing large enough carrier spacing $\Delta f$ and the number of OFDM symbols $N$ in one OTFS frame, as shown in our simulations, the phase compensation matrix can be approximated by all-ones matrix $\mathbf{1}_{M_\tau N \times UM_\tau N}$ while with a little performace degredation. Therefore, (\ref{matrixform}) can be rewritten as
\begin{align}
\label{prob}
	\mathbf Y \approx\ \mathbf X\mathbf H + \mathbf Z.
\end{align}
\begin{figure}
	\centering
	\subfigure[Sparsity due to the user activity with three active users]{
		
		\begin{minipage}{6cm} 
			\includegraphics[width=\textwidth]{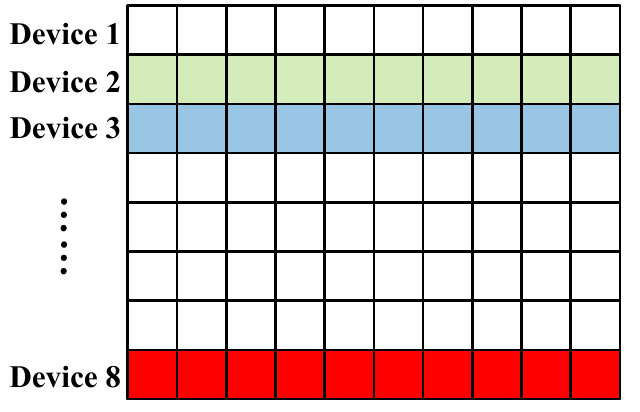} \\
		\end{minipage}
	}
	\subfigure[Sparsity in the angular domain with three active users]{
		\begin{minipage}{5.75cm}%[b]%{0.2\textwidth}
			\includegraphics[width=\textwidth]{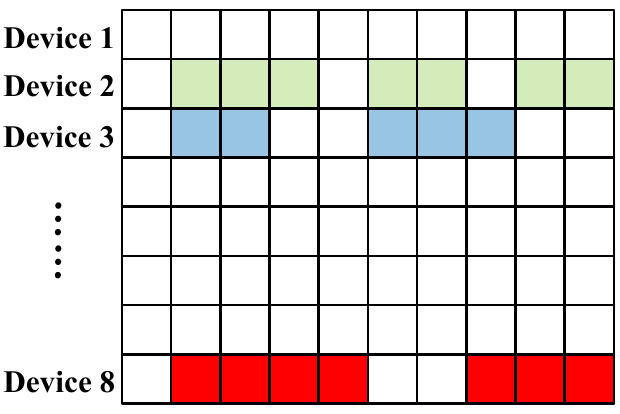} \\
			
		\end{minipage}
	}
	\subfigure[Sparsity in the delay-Doppler-angle domain with two active users]{
	\begin{minipage}{6.2cm}%[b]%{0.2\textwidth}
		\includegraphics[width=\textwidth]{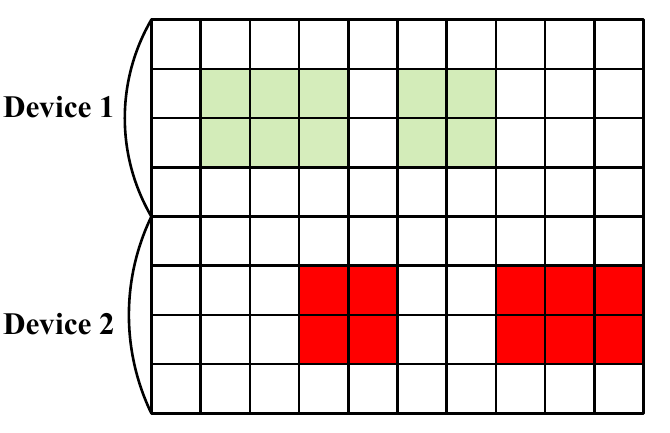} \\
		
	\end{minipage}
}
	\caption{Comparison among different sparsity patterns of the channel matrix.} 
	\label{PAAbefore}
\end{figure}
Note that the channel matrix $\mathbf H$ in the delay-Doppler-angle domian has different kinds of sparsities, compared with that in the previous literature\cite{s31,s32,s33,s34,s43} which only focuses on the sparsity of the user or the sparsity in the angular domain. As shown in Fig. \ref{PAAbefore}, the channel matrix $\mathbf H$ has the 2D burst block sparsity, i.e., not only it has the burst block sparsity\cite{s22} along the column, but also has the burst block sparsity along the row. Specifically, the row sparsity is due to the channel sparsity in the delay-Doppler domain, and the starting positions of the clusters in each column depend on the delay and Doppler shift of all the physical paths of the corrsponding users. To see this, according to  (\ref{angleD}) and (\ref{H_vec}), the ($k+1+lN+uM_\tau N$)-th row of $\mathbf H$ has the dominant elements only if $k \approx NT_{\text{sym}}\nu_{u,i}$ and $l \approx \tau_{u,i} M \Delta f$, and the magnitudes of the elements decrease when $k$ and $l$ move away from $NT_{\text{sym}}\nu_{u,i}$ and $\tau_{u,i} M \Delta f$; the column sparsity of $\mathbf H$ results from the channel sparsity in the angular domain, and the starting positions of the clusters in each row depend on the azimuth angle and elevation angle of the received signal from the corrsponding users, since according to (\ref{angleD}) and (\ref{H_vec}), the $(a_z+1+N_za_y)$-th column of $\mathbf H$ has dominant elements, only if $a_y \approx N_y\Omega_{y_{u}}/2$ and $a_z \approx N_z\Omega_{z_{u}}/2$, and the magnitudes of the elements decrease when $a_y$ and $a_z$ move away from $ N_y\Omega_{y_{u}}/2$ and $N_z\Omega_{z_{u}}/2$. 
%\subsubsection{Device activity sparsity}
%The device activity is sporadic in the grant-free random access systems since only $U_a \ll U$ devices are active in a given time interval. As shown in the Fig[], due to the time-varying channel is considered, each device's channel is represented by $M_\tau N$ rows in $H$, which are much larger than that in the conventional channel matrix[] (e.g., the channel of each device is represented by only one row). This characteristic of $H$ may be benificial to detecting the device activity, since the difference between the channel's energys of the active device and inactive device are larger than that in the conventional channel matrix. Hence, a simple energy-based detection[] with a pre-defined threshold could perform very well when the accurate channel matrix $H$ is acquired.
%\subsubsection{two-dimensional burst block sparsity}

\section{Pattern Coupled SBL Framework}
\label{2}
In this section, a 2D pattern coupled hierarchical prior is first proposed to capture the 2D burst block sparsity in the channel matrix. Then, the SBL framework is developed to estimate the channel matrix using the pilots and the received symbols. Finally, the covariance-free method is adopted to facilitate the computations of the SBL. 
\subsection{The Pattern-Coupled Hierarchical Prior}
Considering that the sparsity
patterns of non-zero coefficients are statistically dependent along the row and column. To explore this 2D sparsity, in the proposed prior, the precision of each coefficient
involves the precisions of its neighborhoods. More precisely, the prior on the $(i,j)$-th element of the channel matrix $\mathbf H$ is given by
\begin{align}
\label{pcprior}
	p(h_{i,j}|\mathrm{\mathbf A},\mathrm{\mathbf B}_{i,j}) = \mathcal{CN}(h_{i,j}|0,\left(\sum_{p=i-D}^{i+D}\sum_{q=j-D}^{j+D} \beta_{i,j,p,q}\alpha_{p,q}\right)^{-1}),
\end{align}
where $\mathrm{\mathbf A}$ is the precision matrix whose $(i,j)$-th element $\alpha_{i,j}$ is the local precision hyperparameter of $h_{i,j}$; $\mathrm{\mathbf B}_{i,j}\in\mathrm{R}^{UM_\tau N\times N_yN_z}$ is the pattern coupled hyperparameter matrix of $h_{i,j}$, and its  $(p,q)$-th element $\beta_{i,j,p,q}$ represents the relevence between $h_{i,j}$ and $h_{p,q}$, and $\beta_{i,j,p,q}\neq0$ only if $p\in[i-D,i+D]$ and $q\in[j-D,j+D]$; $D$, called the range indicatior, is a hyperparameter controlling the number of neighborhoods of $h_{i,j}$ in the consideration. Note that when $\beta_{i,j,p,q} = 0$ for arbitrary $p$ and $q$, except $\beta_{i,j,i,j} = 1$, the proposed prior reduces to the conventional SBL prior\cite{s24}; when $\beta_{i,j,p,q} = 0$ for arbitrary $p$ and $q$, except $\beta_{i,j,i,j-1} = \beta_{i,j,i,j+1} = \zeta$ and $\beta_{i,j,i,j}=1$, where $\zeta$ is a hyperparameter, the proposed prior degrades to
the SBL prior capturing the one dimensional block sparsity\cite{s25}. 

\begin{figure}[!htb]
	\centering
	\includegraphics[width=1.4in]{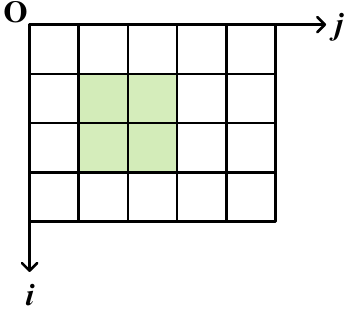}
	\caption{An example of the sparsity pattern of $\mathbf H$ with one user. The non-zero elements are represented by the green boxes, and the other elements are zeros. }
	\label{2D_example}
\end{figure} 
It is clearly seen in (\ref{pcprior}) that the precision of $h_{i,j}$ not only depends on its local hyperparameter $\alpha_{i,j}$, but also on the hyperparameters of its neighborhoods along the row and column of $\mathbf H$, where the relevence between them is controlled by the pattern coupled hyperparameter $\mathrm{\mathbf B}_{i,j}$. It is desirable that $\mathrm{\mathbf B}_{i,j}$ is assigned by the true sparsity pattern of its neighborhoods.
%when any two entries $h_{i,j}$ and $h_{p,q}$ share the same sparsity (i.e., both zeros or non-zeros), their coupled hyparameters $B_{i,j}$ and $B_{p,q}$ 
For example, Fig. \ref{2D_example} shows the sparsity pattern of $\mathbf H$ with one user, where $h_{1,1}, h_{1,2}, h_{2,1}, h_{2,2}$ are non-zero elements.
Suppose $D = 1$ here, and then for $h_{1,1}$, the non-zero pattern can be captured when the coupled hyperparameters  $\beta_{1,1,1,1}=\beta_{1,1,1,2}=\beta_{1,1,2,1}=\beta_{1,1,2,2}=1$, and $\beta_{1,1,0,0}=\beta_{1,1,0,1}=\beta_{1,1,0,2}=\beta_{1,1,1,0}=\beta_{1,1,2,0}=0$. Similarly, for $h_{0,0}$, the zero pattern can be captured when $\beta_{0,0,0,0} = \beta_{0,0,0,1} =\beta_{0,0,1,0} =1$, and $\beta_{0,0,1,1}=0$. Therefore,
%It is observed that the 2D sparsity structure is captured very precisely when the coupled hyperparameter $B_{i,j}$ of non-zero elements can be assigned by the support of $H$. 
such coupled prior has the potential to capture the unknown 2D sparsity structure in $\mathbf H$, where the pattern coupled hyperparameter $\mathrm{\mathbf B}_{i,j}$ plays an important role, and we will discuss the update of it later. Combined with (\ref{pcprior}), the  prior distribution of $\mathbf H$ is given by
\begin{align}
\label{pH}
p(\mathbf H|\mathrm{\mathbf A},\mathrm{\mathbf B}) &=\prod_{i=0}^{UM_\tau N-1} \prod_{j=0}^{N_yN_z-1} p\left(h_{i,j} | \mathrm{\mathbf A},\mathrm{\mathbf B}_{i,j}\right) \nonumber\\
&=  \prod_{j=0}^{N_yN_z-1} \mathcal{CN}(\mathbf{h}_j| 0, \Upsilon_j),
\end{align}
where $\mathrm{\mathbf B}$ contains all the pattern coupled hyperparameters, whose $(i,j)$-th element is $\mathrm \mathrm{\mathbf B}_{i,j}$;  $\mathbf{h}_j=[h_{0,j},\dots,h_{UM_\tau N -1,j}]$ and represents the $j$-th column of $\mathbf H$; $\Upsilon_j = \mathrm{diag}(\gamma_{0,j},\dots,\gamma_{UM_\tau N -1,j})$ is the covariance matrix of $\mathbf{h}_j$, and $\gamma_{i,j} = \left(\sum_{p=i-D}^{i+D}\sum_{q=j-D}^{j+D} \beta_{i,j,p,q}\alpha_{p,q}\right)^{-1}$. Then, following the conventional SBL, we adopt the Gamma distributions as the hierarchical prior for the local precision hyperparameter $\mathrm{\mathbf A}$, i.e.,

\begin{align}
	p(\mathrm{\mathbf A}|a,b) = \prod_{i=0}^{UM_\tau N-1} \prod_{j=0}^{N_yN_z-1} \Gamma(a)^{-1} b^{a} \alpha_{i,j}^{a} e^{-b \alpha_{i,j}},
\end{align}
where $\Gamma(c)\triangleq\int_{0}^{\infty} t^{c-1} e^{-t} d t$. This hierarchical Beysian modeling provides the analytically tractable solutions for the estimation of $\mathbf H$, and encourages the sparseness in the estimation since the overall prior of $\mathbf H$ (i.e., integrate
out $\alpha_{i,j}$) is a Student-$t$ distribution which is sharply peaked at zero. In the conventional SBL, $a$ and $b$ are usually assigned by the small values (e.g., $10^{-4}$), called the non-informative prior\cite{s24}. In this paper, we adopt a informative prior and set $a = 1$ since as we stated above, the channel in the delay-Doppler-angle domain is very sparse and the larger $a$ will promote the sparseness in the estimation\cite{s25}.
\subsection{SBL Framework}
\subsubsection{MMSE Estimation of $\mathbf H$} We now develop the SBL framework for the 2D burst block sparse signal recovery. Recalling that the noise obey a Guassian distribution with the precision $\theta$, and then the likelihood function is given by
\begin{align}
\label{likeli}
	p(\mathbf Y|\mathbf H,\theta)=\prod_{j=0}^{N_yN_z-1} \mathcal{CN}(\mathbf{y}_j| \mathbf X\mathbf{h}_j, \theta^{-1}\mathbf I_{UM_\tau N}),
\end{align}
where $\mathbf{y}_j=[y_{0,j},\dots,y_{UM_\tau N-1}]$ is the $j$-th column of $\mathbf Y$. According to the Bayes’ rule,  the posterior distribution of $\mathbf H$ is given as
\begin{align}
\label{poster1}
	p(\mathbf H|\mathbf Y,\mathrm{\mathbf A},\mathrm{ \mathbf B},\theta) = \frac{p(\mathbf Y|\mathbf H,\theta)p(\mathbf H|\mathrm{\mathbf A},\mathrm{\mathbf B})}{\int p(\mathbf Y|\mathbf H,\theta)p(\mathbf H|\mathrm{\mathbf A},\mathrm {\mathbf B}) d\mathbf H }.
\end{align}
Since in (\ref{poster1}), the distributions appearing in the numerator are Gaussians, and the normalising integral is a convolution of Gaussians. Thus, the posterior distribution of $\mathbf H$ is also Gaussian, and is given by
\begin{align}
	p(\mathbf H|\mathbf Y,\mathrm{\mathbf A},\mathrm{\mathbf B},\theta) = \prod_{j=0}^{N_yN_z-1} \mathcal{CN}(\mathbf{h}_j|\mu_j, \mathbf{\Sigma}_j),
\end{align}
where 
\begin{align}
\label{esti_mu}
	\mu_j = \theta\mathbf{\Sigma}_j\mathbf X^{\mathrm{H}}\mathbf{y}_j, \mathbf{\Sigma}_j = (\theta \mathbf X^{\mathrm{H}}\mathbf X+\Upsilon_j^{-1})^{-1}.
\end{align}
Given a set of estimated hyperparameters $\{\mathrm{ \mathbf A},\mathrm{\mathbf B},\theta\}$, the MMSE estimation of $\hat{\mathbf H}$ is the mean of its posterior distribution, i.e.,
\begin{align}
	\hat{\mathbf H} = [\mu_0,\dots,\mu_{N_yN_z-1}].
\end{align}
Then, our problem reduces to estimating the set of hyperparameters.
\subsubsection{Learning the Local Hyperparameters} Follwing the conventional SBL, we place a Gamma hyperpriors over $\theta$, i.e.,
\begin{align}
	p(\theta|r,s) = \Gamma(r)^{-1} s^{r} \theta^{r} e^{-s \theta}.
\end{align}
Then, the precision matrix $\mathrm{\mathbf A}$ and the noise precision $\theta$ can be estimated by maximizing their log-posterior distribution with respect to $\mathrm{\mathbf A}$ and $\theta$, i.e.,

\begin{align}
\label{max_p}
	\hat{\mathrm{\mathbf A}},\hat{\theta}=\arg \max _{\mathrm{\mathbf A},\theta} \log p(\mathrm{\mathbf A},\theta |\mathbf Y, \mathrm{\mathbf B}),
\end{align}
where the pattern coupled hyperprameter $\mathrm{\mathbf B}$ is assumed to be known in this step. 

Next, we adopt the EM algorithm to approximate the optimal solutions in (\ref{max_p}), where the channel matrix $\mathbf H$ is treated as the hidden variables. Given the estimations of the hyperparameters $\mathrm{\mathbf A}^t$, $\theta^t$ in the $t$-th iteration, and the received symbolds $\mathbf Y$, we denote $\mathrm{E}_{\mathbf H|\mathbf Y,\mathrm{\mathbf A}^t,\theta^t}[\cdot]$ as $\mathrm{E}[\cdot]$ for notation simplicity, where the expection is with respect to the channel $\mathbf H$ over the distribution $p(\mathbf H|\mathbf Y,\mathrm{\mathbf A}^t,\theta^t)$. Then, the Q-function is given by
\begin{align}
\label{Q_func}
Q(\mathrm{\mathbf A},&\theta|\mathrm{\mathbf A}^t,\theta^t) \nonumber \\
= &\mathrm{E}[\log p(\mathbf Y,\mathbf H,\mathrm{\mathbf A},\theta|\mathrm{\mathbf B})] \nonumber \\
=&\mathrm{E}[\log p(\mathbf Y|\mathbf H,\theta)p(\theta)] + \mathrm{E}[\log p(\mathbf H|\mathrm{\mathbf A},\mathrm{\mathbf B})p(\mathrm{\mathbf A})].	
\end{align}
It is observed in (\ref{Q_func}) that the parameters $\mathrm{\mathbf A}$ and $\theta$ are seperated from each other. This allows the estimations of $\mathrm{\mathbf A}$ and $\theta$ to be decoupled into the two independent problems, i.e.,
\begin{align}
	&\mathrm{\mathbf A}^{t+1}=\arg \max _{\mathrm{A}} \mathrm{E}[\log p(\mathbf H|\mathrm{\mathbf A},\mathrm{\mathbf B})p(\mathrm{\mathbf A})]\label{Am},\\
	&\theta^{t+1} = \arg \max _{\theta} \mathrm{E}[\log p(\mathbf Y|\mathbf H,\theta)p(\theta)]\label{noisep}.
\end{align}
The optimization of (\ref{Am}) in the conventional SBL can be decoupled into a number of separate
problems in which each hyperparameter is updated independently\cite{s24}, and usually, the analytic solution is easy to obtain.
However, it is not the case for the problem being
considered here. To see this, (\ref{Am}) can be written as
\begin{align}
\label{A_f}
&\mathrm{\mathbf A}^{t+1}=\arg \max _{\mathrm{A}} \sum_{i} \sum_{j} \log \left(\sum_{p=i-D}^{i+D}\sum_{q=j-D}^{j+D} \beta_{i,j,p,q}\alpha_{p,q}\right)\nonumber\\ &-\left(\sum_{p=i-D}^{i+D}\sum_{q=j-D}^{j+D} \beta_{i,j,p,q}\alpha_{p,q}\right)\mathrm{E}[\left|h_{i j}\right|^{2}]+a \log \alpha_{i, j}-b \alpha_{i, j},
\end{align}
where the hyperparameters in (\ref{A_f}) are entangled with each other due to the logarithm term. Therefore, the analytical solution of the optimization is difficult to obtain. Besides, although the gradient descent
methods can be adopted to get the optimal solution, 
it involves higher computational complexity as compared with an analytical update rule. Similar to \cite{s24}, we derive a closed form sub-optimal solution by examining the optimality in (\ref{A_f}), as shown in the following proposition.

\textbf{Proposition 2:} The range of the optimal $(i,j)$-th element $\alpha_{i, j}^*$ in the $\mathrm{\mathbf A^{t+1}}$ is given by
\begin{align}
	\left[\frac{a}{b+\omega_{i,j}^{t}},\frac{a+(2D+1)^2}{b+\omega_{i,j}^{t}}\right],
\end{align}
where $\omega_{i,j}^{t}=\sum_{p=i-D}^{i+D}\sum_{q=j-D}^{j+D}\beta_{p,q,i,j}\mathrm{E}[\left|h_{p q}\right|^{2}]$.

\emph{Proof}: See Appendix \ref{p2}.

Therefore, a sub-optimal solution to (\ref{Am}) can  be simply chosen as
\begin{align}
\label{alpha}
	\alpha_{i,j}^{t+1}=\hat \alpha_{i, j} = \frac{a}{b+\omega_{i,j}^{t}}.
\end{align}
Finally, the second moment $\mathrm{E}[\left|h_{p q}\right|^{2}]=\left|\mu^t_{p,q}\right|^2 + \left|\Sigma^t_{p,q}\right|^2$, where $\mu^t_{p,q}$ is the $p$-th element of the posterior mean $\mu^t_q$ in the $t$-th iteration, and $\Sigma^t_{p,q}$ is the $(p,p)$-th entry of the covariance matrix $\mathbf{\Sigma}^t_q$.

For the estimation of $\theta$, (\ref{noisep}) can be written as 
\begin{align}
\label{thetap}
	\theta^{t+1} = \arg \max _{\theta} \sum_{j=0}^{N_yN_z-1}\left(MN\log\theta-\theta\mathrm{E}\left[\left\|\mathbf{y}_j-\mathbf X\mathbf{h}_j\right\|^2\right]\right) \nonumber \\
	+ r\log\theta - s\theta.
\end{align}
Computing the first derivative of the objective function in (\ref{thetap}) with respect to $\theta$ and setting it equal to zero, we get
\begin{align}
\label{theta_up}
	\theta^{t+1}=\hat \theta = \frac{M_\tau N N_yN_z+r}{\sum_{j=0}^{N_yN_z-1}\mathrm{E}\left[\left\|\mathbf{y}_j-\mathbf X\mathbf{h}_j\right\|^2\right] +s },
\end{align}
where
\begin{align}
\mathrm{E}\left[\left\|\mathbf{y}_j-\mathbf X\mathbf{h}_j\right\|^2\right] = \left\|\mathbf{y}_j-\mathbf X\mu^t_j\right\|^2 
+ (\theta^t)^{-1}\nonumber\\\times\sum_{i=0}^{UM_\tau N-1}(1-\Sigma^t_{i,j}(\gamma^t_{i,j})^{-1}),
\end{align}
and $\gamma^t_{i,j}$ is the prior variance of $h_{i,j}$ in the $t$-th iteration.
\subsubsection{Updating the Coupled Hyperparameters}
As discussed in the last subsection, the coupled hyperparameter matrices $\mathrm{\mathbf B}_{i,j}$ of $h_{i,j}$ can be assigned by the sparsity pattern of its neighborhoods. Here, we will first estimate the support of $\mathbf H$ based on the current estimations of hyperparameters and then, use it to update $\mathrm{\mathbf B}$. Consider the following binary hypothesis testing
\begin{align}
	\begin{array}{ll}
	\mathcal{H}_{0}: & \gamma_{i,j}=0 \\
	\mathcal{H}_{1}: & \gamma_{i,j}>0
	\end{array},
\end{align}
where $\mathcal{H}_{0}(\mathcal{H}_{1})$ represents that $h_{i,j}$ is zero (non-zero). Similar to \cite{ss1,ss2}, we consider the log-likelihood ratio (LLR) test, given by
\begin{align}
\label{oLLR}
	\log \frac{p\left(\mathbf{y}_j | \mathcal{H}_{1}\right)}{p\left(\mathbf{y}_j | \mathcal{H}_{0}\right)} \geqslant \xi,
\end{align}
where $\xi$ is the detection threshold. Similar to the denominator in (\ref{poster1}), $p\left(\mathbf{y}_j | \mathcal{H}_{1}\right)$ as well as $p\left(\mathbf{y}_j | \mathcal{H}_{0}\right)$ can be written as the convolution of two Gaussians, and thus is given by
\begin{align}
	p\left(\mathbf{y}_j | \mathcal{H}_{1}\right) = \mathcal{CN}(0,\theta^{-1}\mathbf I_{M_\tau N} + \mathbf X\Upsilon_j\mathbf X^{\mathrm{H}}), \\
	p\left(\mathbf{y}_j | \mathcal{H}_{0}\right) = \mathcal{CN}(0,\theta^{-1}\mathbf I_{M_\tau N} + \mathbf X\Upsilon^{\prime}_j\mathbf X^{\mathrm{H}}), 
\end{align}
where $\Upsilon^{\prime}_j = [\gamma_{0,j},\dots,\gamma_{i-1,j},0,\gamma_{i+1,j},\dots,\gamma_{UM_\tau N-1,j}]$. Then, the LLR test in (\ref{oLLR}) is equivalent to
\begin{align}
\label{eql}
	\frac{\left|\mathbf{x}_{i}^{\mathrm H} 
		\left(\theta^{-1} I_{M_\tau N}+\mathbf X\Upsilon_j^{\prime} \mathbf X^{\mathrm H}\right)^{-1} \mathbf{y}_j\right|^{2}}{\mathbf{x}_{i}^{\mathrm H} 
		\left(\theta^{-1} I_{M_\tau N}+\mathbf X\Upsilon_j^{\prime} \mathbf X^{\mathrm H}\right)^{-1}\mathbf x_i} \geqslant \bar{\xi}.
\end{align}
According to the Neyman-Pearson lemma\cite{sd}, the threshold $\bar{\xi} = \log \frac{1}{P_{fa}}$, and $P_{fa}$ is false alarm probability, since the random variable on the left side of the inequality obeys the chi-square distribution with two degrees of freedom under the hypothesis $\mathcal{H}_{0}$. We denote the support of $\mathbf H$ as $\mathbf S_H$, and then according to (\ref{eql}), the $(i,j)$-th element of the $\mathbf S_H$ is given by
\begin{align}
	S_{H_{i,j}} = \mathbb{I}\left\{\frac{\left|\mathbf{x}_{i}^{\mathrm H} 
			\left(\theta^{-1}\mathbf I_{M_\tau N}+\mathbf X\Upsilon_j^{\prime} \mathbf X^{\mathrm H}\right)^{-1} \mathbf{y}_j\right|^{2}}{\mathbf{x}_{i}^{\mathrm H} 
			\left(\theta^{-1}\mathbf I_{M_\tau N}+\mathbf X\Upsilon_j^{\prime} \mathbf X^{\mathrm H}\right)^{-1}\mathbf x_i} \geqslant \log \frac{1}{P_{fa}}\right\}
\end{align}
where $i=0,\dots,UM_\tau N-1$, $j=0,\dots,N_yN_z-1$, and $\mathbb{I}\{\cdot\}$ is the indicatior function which is 1 if the condition inside the braces is fulfilled and 0 otherwise. Based on $\mathbf S_H$, the pattern coupled hyperparameter $\mathbf B_{i,j}$ can be updated by the support of the neighborhoods of $h_{i,j}$. For $p\in[i-D,i+D]$ and $q\in[j-D,j+D]$, $\beta_{i, j, p,q}$ is given by
\begin{align}
\label{beta_update}
	\beta_{i, j, p,q}= \begin{cases}S_{H_{p,q}}, & \text { if } S_{H_{i,j}=1}, \\ 1 -  S_{H_{p,q}}, & \text { otherwise. }\end{cases}
\end{align}
Finally, based on the estimation of the support of $\mathbf H$, the device activity detection rule is given by
\begin{align}
\label{lab}
	\hat{\lambda}_u = \mathbb{I}\left\{\sum_{i=uM_\tau N}^{(u+1)M_\tau N-1}\sum_{j=0}^{N_yN_z-1} S_{H_{i,j}}>0\right\}. 
\end{align}
\subsubsection{The Covariance-free Method}
Indeed, the above method can effectively detect the device activity and estimate the channel. However, to compute the covariance matrix in (\ref{esti_mu}) and (\ref{eql}), the costly matrix inversions are involved, which limits its application for the large-scale systems. Here, we adopt the covariance-free method\cite{s28} to facilitate the computations of the above SBL algorithm. The main idea is that transforming the matrix inversions into solving the linear equations which can be computed faster by the effective iterative algorithms. To see this, we re-express (\ref{esti_mu}) and the terms in (\ref{eql}) involved the matrix inversions as
\begin{align}
	&(\theta \mathbf X^{\mathrm{H}}\mathbf X+\Upsilon_j^{-1}) \mu_j = \theta \mathbf X^{\mathrm{H}}\mathbf{y}_j  \label{muesti}\\
	&\left(\theta^{-1} I_{M_\tau N}+\mathbf X\Upsilon_j^{\prime} \mathbf X^{\mathrm H}\right)\mathbf{t}_{nu} = \mathbf{y}_j \label{test1}  \\
	&\left(\theta^{-1} I_{M_\tau N}+\mathbf X\Upsilon_j^{\prime} \mathbf X^{\mathrm H}\right)\mathbf{t}_{de} = \mathbf{x}_i \label{test2}. 
\end{align}
Therefore, the posterior mean $\mu_j$ and the terms in (\ref{eql}) involved the matrix inversions are the solutions of the above linear equations, where $\mathbf{t}_{nu} = \left(\theta^{-1} I_{M_\tau N}+\mathbf X\Upsilon_j^{\prime} \mathbf X^{\mathrm H}\right)^{-1} \mathbf{y}_j$ and $\mathbf{t}_{de} = \left(\theta^{-1} I_{M_\tau N}+\mathbf X\Upsilon_j^{\prime} \mathbf X^{\mathrm H}\right)^{-1} \mathbf{x}_i$. Note that except for the posterior mean, the update of the hyperparameter still need the posterior variance of the channel $h_{i,j}$ (i.e., the diagonal elements of the covariance matrix $\Sigma_j$), which can be estimated by the following proposition\cite{math}.

\textbf{Proposition 3}: Let $\mathrm{\mathbf S}$ be any square matrix of size $W \times W .$ Let $\mathbf p_{1}, \mathbf p_{2}, \ldots, \mathbf p_{K} \in \mathrm{R}^{W}$ be $K$ random probe vectors, where each $\mathbf p_{k}$ is comprised of independent and identically distributed components such that $\mathrm{E}\left[\mathbf p_{k}\right]=\mathbf 0$. Consider the estimator
$$
\mathbf s_d=\left(\sum_{k=1}^{K} \mathbf p_{k} \odot \mathrm{\mathbf S} \mathbf p_{k}\right) \oslash\left(\sum_{k=1}^{K} \mathbf p_{k} \odot \mathbf p_{k}\right),
$$
where $\oslash$ denotes element-wise division between two vectors. Then, the vector $\mathbf s_d$ is an unbiased estimator of the diagonal elements of $\mathrm{\mathbf S}$. 

We apply this diagonal estimation rule to $\mathbf{\Sigma}_j$ to estimate its diagonal elements vector $\mathbf s_{d_j}$. The simplest distribution to use in drawing probe vectors $\mathbf p_{k}$ is the Randemacher distribution, which lets each independent component of $\mathbf p_{k}$ be either $-1$ or $+1$ with equal probability. In this case, the diagonal estimator $\mathbf s_{d_j}$ simplifies to
\begin{align}
\label{sigesti}
	\mathbf s_{d_j}=\frac{1}{K} \sum_{k=1}^{K} \mathbf p_{k} \odot \mathbf{\Sigma}_j \mathbf p_{k}.
\end{align}
Similarly, the term $\mathbf{\Sigma}_j \mathbf p_{k}$ can be re-expressed as 
\begin{align}
\label{sigliea}
	(\theta \mathbf X^{\mathrm{H}}\mathbf X+\Upsilon_j^{-1})\mathbf{t}_{d,j} = \mathbf p_{k},
\end{align}
where $\mathbf{t}_{d,j} = \mathbf{\Sigma}_j \mathbf p_{k}$. Until now, all the needed computations involved the matrix inversions have been transformed into solving the linear equations, and then the conjugate gradient (CG) algorithm\cite{cg} can be adopted to tackle these linear equations.

The TDSBL-CF\footnote{For the LEO satellite communications, the complexity of the TDSBL-CF is acceptable when the number of potential devices is small or moderate, and the grant-free mechanism offers the possibility of low access latency.}
is summarized in Algorithm \ref{TDSBL-FM}. Note that in the Algorithm \ref{TDSBL-FM}, we collect all the activity indicators $\lambda_u$ into a single vector $\lambda = [\lambda_0,\dots,\lambda_{U-1}]$, and the algorithm will output its estimation $\hat{\lambda}$. The complexity of TDSBL-CF mainly depends on the complexity of iterative algorithms solving the linear equations. For example, when the CG algorithm is adopted, the complexity is $\mathcal{O}((K+1)N_yN_z(UM_\tau N)^2+ 2(M_\tau N)^3UN_yN_z)$ in each iteration, where the quadratic term is due to the CG algorithm for estimating the posterior mean and variance, and the cubic term is due to the LLR test for each element in the channel matrix. Compared with the complexity $\mathcal{O}(N_yN_z(UM_\tau N)^3+ 2(M_\tau N)^4UN_yN_z) $ of using the matrix inversions directly, the TDSBL-CF has greatly reduced the computations. However, it is still prohibitive for the very large-scale systems, since typically the number of carriers adopted in the OTFS and the potential devices are huge. 
To facilitate the TDSBL-CF for solving the large-scale problems, the GAMP-based algorithm is developed in the next section. 
%its computations will increase sharply when the large number of carriers is adopted and more potential devices are intended to access the satellite. 
\begin{algorithm}[hbt!]
	\caption{TDSBL-CF Algorithm}\label{TDSBL-FM}
	\begin{algorithmic}[1]
		\REQUIRE{Recieved symbols $\mathbf Y$, pilot matrix $\mathbf X$; the maximum number of iterations $T_{SBL}$, the termination threshold $\eta$, the false alarm probability $P_{fa}$, the range indicatior $D$, and the number of probes $K$.}
		\ENSURE The estimated channel matrix $\hat{\mathbf H}$ and the activity indicatior vector $\hat{\lambda}$.
		\STATE Initialization: $\forall i,j$: $\mu_{i,j}^0=0$, $\mu_{i,j}^1=10^{-8}$ $\gamma_{i,j}^0=10^{-2}$, $\beta^0_{i,j,p,q} = 1$ for $p \in [i-D,i+D]$, $q \in [j-D,j+D]$, and 0 otherwise; $\theta^0=10^{3}$, and $t = 0$.
		\WHILE{$t < T_{SBL}$ $\&$ $\sum_{j=0}^{N_yN_z-1}\frac{\left\|\mu^t_j-\mu^{t+1}_j\right\|^2}{\left\|\mu^{t}_j\right\|^2} > \eta$}
		\STATE $t \gets t+1$
		\STATE $\forall i,j$: Update posterior mean $\mu^t_{i,j}$ and variance $\Sigma_{i,j}^t$ by solving the linear equations in (\ref{muesti}), (\ref{sigesti}) and (\ref{sigliea}).
		\STATE Update the noise precision $\theta^t$ according to (\ref{theta_up}).
		\STATE $\forall i,j$: Update the local precision parameter $\alpha_{i, j}$ according to (\ref{alpha}).
		\STATE $\forall i,j$: Update the support matrix $S^t_{H_{i,j}}$ according to (\ref{eql}) and by solving the linear equations (\ref{test1}) and (\ref{test2}).
		\STATE $\forall i,j$: Update $\beta_{i,j,p,q}^t$ according to (\ref{beta_update})
		\STATE $\forall u$: Update $\lambda_u^t$ according to (\ref{lab}).
		%\IF{$N$ is even}
		%\STATE $X \gets X \times X$
		%\STATE $N \gets \frac{N}{2} $  \COMMENT{This is a comment}
		%\ELSIF{$N$ is odd}
		%\STATE $y \gets y \times X$
		%\STATE $N \gets N - 1$
		%\ENDIF
		\ENDWHILE
		\RETURN $\hat{\mathbf H}=[\mu_0^t,\dots,\mu_{N_yN_z-1}^t]$ and $\hat{\lambda}=[\lambda^t_0,\dots,\lambda^t_{U-1}]$
	\end{algorithmic}
\end{algorithm}

%Overall, the two-dimensional pattern coupled SBL with the covariance free method (TDSBL-CF) is summarized in algorithm 1.

\section{Low Complexity SBL with GAMP}
\label{3}
In this section, we further decrease the complexity of the above SBL framework by adopting the GAMP algorithm to estimate the posterior mean and variance of the channel matrix, and by considering a specical case of the proposed prior without updating the coupled hyperparameters. Instead of approximating the solution of (\ref{esti_mu}), the GAMP algorithm directly estimates $\mu_{i,j}$ and $\Sigma_{i,j}$ by utilizing the factorization of the prior and likelihood function of the channel matrix.
%To further decrease the complexity of the above SBL method,
  
Recalling that the prior of $h_{i,j}$ in (\ref{pH}) and the likelihood function in (\ref{likeli}) can be written as the product of multiple probability density functions, and then the posterior distribution of $h_{i,j}$ is given by
\begin{align}
p( h_{i,j}&|\mathbf Y,\mathrm{\mathbf A},\mathrm{\mathbf B},\theta) \propto \int p(\mathbf y_j|\mathbf g_j,\theta)p(\mathbf h_j|\mathrm{\mathbf A}, \mathrm{\mathbf B}) dh_{\sim i,j}\nonumber\\
= &\int\prod_{o=0}^{M_\tau N-1}p(y_{o,j}|g_{o,j},\theta) \prod_{i=0}^{UM_\tau N-1}p(h_{i,j}|\mathrm{\mathbf A},\mathrm{\mathbf B})dh_{\sim i,j},
\end{align}
where $\int (\cdot) dh_{\sim i,j}$ represents the integral with respect to $\mathbf h_j$ except $h_{i,j}$, $\mathbf g_j=\mathbf X\mathbf h_j$, and $g_{o,j}$ is the $o$-th element of $\mathbf g_j$. This implies that the message passing algorithms can be adopted to estimate the posterior mean and variance of $h_{i,j}$, instead of directly computing the matrix inversion in (\ref{esti_mu}), which will deacrease the computational complexity sharply. Here, similar to \cite{s42}, we resort to the GAMP algorithm. According to \cite{sss}, based on the estimations of all the hyperparameters in the $t$-th iteration, the approximations of the posterior mean and variance of $h_{i,j}$ in $(t+1)$-th iteration is given by
\begin{align}
	\mu_{i,j}^{t+1} =  \mathrm{E}[h_{i,j}|\hat{r}_{i,j}^t,\tau^{r,t}_{i,j},\mathrm{\mathbf A},\mathrm{\mathbf B}],\\
	 \Sigma_{i,j}^{t+1}=\mathrm{Var}[h_{i,j}|\hat{r}_{i,j}^t,\tau^{r,t}_{i,j},\mathrm{\mathbf A},\mathrm{\mathbf B}],
\end{align} 
where the expectation is with respect to $p(h_{i,j}|\hat{r}_{i,j}^t,\tau^{r,t}_{i,j},\mathrm{\mathbf A},\mathrm{\mathbf B})$, which is related to the prior of $h_{i,j}$ and given by
\begin{align}
	p(h_{i,j}|\hat{r}_{i,j}^t,&\tau^{r,t}_{i,j},\mathrm{\mathbf A},\mathrm{\mathbf B})\nonumber \\&=
	\frac{p(h_{i,j}|\mathrm{\mathbf A}, \mathrm{\mathbf B})\mathcal{CN}(h_{i,j}|\hat{r}_{i,j}^t,\tau^{r,t}_{i,j})}{\int p(h_{i,j}|\mathrm{\mathbf A}, \mathrm{\mathbf B})\mathcal{CN}(h_{i,j}|\hat{r}_{i,j}^t,\tau^{r,t}_{i,j})dh_{i,j}},
\end{align}
Since the prior of $h_{i,j}$ is Guassian, and thus $p(h_{i,j}|\hat{r}_{i,j}^t,\tau^{r,t}_{i,j},\mathrm{\mathbf A},\mathrm{\mathbf B})$ is also Gaussian. Then, $ \mu_{i,j}^{t+1}$ and $ \Sigma_{i,j}^{t+1}$ can be easily computed as
\begin{align}
\label{mmm}
	\mu_{i,j}^{t+1} = \frac{\gamma_{i,j}^t\hat{r}_{i,j}^t}{\gamma_{i,j}^t+\tau^{r,t}_{i,j}},  \\
	\Sigma_{i,j}^{t+1} = \frac{\gamma_{i,j}^t\tau^{r,t}_{i,j}}{\gamma_{i,j}^t+\tau^{r,t}_{i,j}},\label{mmm2}
\end{align}
where 
\begin{align}
\label{ee}
	&\tau^{r,t}_{i,j}=\left(\sum_{o} \left|X_{o,i} \right|^2 \tau_{o,j}^{s,t}\right)^{-1},\\
	%\left(1-\frac{1}{\tau_{o,j}^{p,t}} %\mathrm{Var}\left[g_{o,j} | y_{o,j}, \hat{p}_{o,j}^t,\tau_{o,j}^{p,t}\right]\right)\right)^{-1}\\
	&\hat{r}_{i,j}^t=\mu_{i,j}^t+\tau^{r,t}_{i,j} \sum_{o} X_{o, i}^{*} \hat{s}_{o,j}^t, \label{ee1}
	% \left(\mathrm E\left[g_{o,j} | y_{o,j}, \hat{p}_{o,j}^t,\tau_{o,j}^{p,t}\right]-\hat{p}_{o,j}^t\right)\label{vv}
\end{align}
and
\begin{align}
\label{ss1}
	&\hat{s}_{o,j}^t = \frac{1}{\tau_{o,j}^{p,t}} \left(\mathrm E\left[g_{o,j} | y_{o,j}, \hat{p}_{o,j}^t,\tau_{o,j}^{p,t}\right]-\hat{p}_{o,j}^t\right),\\
	&\tau_{o,j}^{s,t} = \frac{1}{\tau_{o,j}^{p,t}}(1-\frac{1}{\tau_{o,j}^{p,t}}\mathrm{Var}\left[g_{o,j} | y_{o,j}, \hat{p}_{o,j}^t,\tau_{o,j}^{p,t}\right]).
\end{align}
The expectation in (\ref{ss1}) is with respect to $p(g_{o,j} | y_{o,j}, \hat{p}_{o,j}^t,\tau_{o,j}^{p,t})$, which is related to the likelihood function and is given by
\begin{align}
	p(g_{o,j} | y_{o,j}, \hat{p}_{o,j}^t,&\tau_{o,j}^{p,t})\nonumber \\
	=&\frac{p(y_{o,j}|g_{o,j})\mathcal{CN}(g_{o,j}|\hat{p}_{o,j}^t,\tau_{o,j}^{p,t})}{\int p(y_{o,j}|g_{o,j})\mathcal{CN}(g_{o,j}|\hat{p}_{o,j}^t,\tau_{o,j}^{p,t})dg_{o,j}},
\end{align}
Note that the likelihood is also Gaussian, and so is 	$p(g_{o,j} | y_{o,j}, \hat{p}_{o,j}^t,\tau_{o,j}^{p,t})$. Then, we can get
\begin{align}
	&\mathrm E\left[g_{o,j} | y_{o,j}, \hat{p}_{o,j}^t,\tau_{o,j}^{p,t}\right] = \hat{p}_{o,j}^t + \frac{\tau_{o,j}^{p,t}}{\tau_{o,j}^{p,t}+(\theta^t)^{-1}}(y_{o,j}-\hat{p}_{o,j}^t), \\
	&\mathrm{Var}\left[g_{o,j} | y_{o,j}, \hat{p}_{o,j}^t,\tau_{o,j}^{p,t}\right]=\frac{\theta^{-1}\tau_{o,j}^{p,t}}{(\theta^t)^{-1}+\tau_{o,j}^{p,t}},
\end{align}
where
\begin{align}
	&\tau_{o,j}^{p,t}=\sum_{i}\left|X_{o,i}\right|^2 \Sigma_{i,j}^t,\\
	&\hat{p}_{o,j}^t = \sum_{i} (X_{o,i}\hat \mu_{i,j}^t) -
	\hat{s}_{o,j}^{t-1}\tau_{o,j}^{p,t} \label{pp},
	 %\frac{\tau_{o,j}^{p,t}}{\tau_{o,j}^{p,t-1} + (\theta^{t-1})^{-1}}(y_{o,j}-\hat{p}_{o,j}^{t-1})
\end{align}

Then, to further decrease the complexity of the proposed SBL method, we consider a special case of the proposed prior in (\ref{pcprior}), where we set the coupled parameter $\beta_{i,j,i,j}=1$, and $\beta_{i,j,p,q} = \beta$ for $p\in[i-D,i+D]$ and $q\in[j-D,j+D]$. Then, (\ref{pcprior}) is simplified as
\begin{align}
\label{newp}
&p(h_{i,j}|\mathrm{\mathbf A},\mathrm{\mathbf B}_{i,j})\nonumber\\ &= \mathcal{CN}(h_{i,j}|0,\left(\alpha_{i, j} +\beta(\sum_{p=i-D,p\neq i}^{i+D}\sum_{q=j-D,q\neq j}^{j+D}\alpha_{p,q})\right)^{-1}),
\end{align}
Here, the single parameter $\beta$ is adopted to represent the relevence between the $h_{i,j}$ and its neighborhoods, and will not be updated using the sparsity pattern of $\mathbf H$ for decreasing the computations.
Then, the variance of $h_{i,j}$ in the $t$-th iteration can be computed by the 2D convolution between the local precison matrix $\mathrm{\mathbf A}^t$ and a kernel $\mathcal{B}$, i.e.,
\begin{align}
\label{ggg}
(\gamma_{i,j}^t)^{-1} = \sum_{m=-\infty}^{\infty}\sum_{n=-\infty}^{\infty} \mathrm{\mathbf A}^t_{m,n} \mathcal{B}_{i-m,j-n}, 
\end{align}
where $\mathcal{B}\in\mathrm R^{(2D+1)\times(2D+1)}$ and only the $(D,D)$-th entry of $\mathcal{B}$ is equal to one, and other entries are $\beta$. 
%For example, $\mathcal{B}$ is given as
%\begin{align}
	%\mathcal{B}=\left[\begin{array}{lll}
	%\beta &\beta  &\beta \\
	%\beta &1  &\beta \\
	%\beta &\beta &\beta
	%\end{array}\right],
%\end{align} 
%if we set $D=1$. 
Besides, according to (\ref{alpha}), $\alpha_{i,j}^{t+1}$ is given by
\begin{align}
\label{aaa}
	\alpha_{i,j}^{t+1} = \frac{a}{b+\omega^{t+1}},
\end{align}  
where $\omega^{t+1}$ is the 2D convolution between the kernel $\mathcal{B}$ and the second moment matrix $\Psi^t$ whose $(i,j)$-th element is $\left|\mu^t_{i,j}\right|^2 + \left|\Sigma^t_{i,j}\right|^2$, and then is given by
\begin{align}
\label{aaa1}
	\omega^{t+1} = \sum_{m=-\infty}^{\infty}\sum_{n=-\infty}^{\infty} \mathrm \Psi^t_{m,n} \mathcal{B}_{i-m,j-n},
\end{align}
Note that the special prior leads to the 2D-convolution-based update in the EM algorithm, considering the elements both along the delay-Doppler domain and angular domain, and thus matches the 2D burst block sparsity. Finally, the precision of noise can be updated according to (\ref{theta_up}). When the channel has been estimated, we propose to adopt the energy-based detector to get the active device, i.e.,
\begin{align}
\label{ddd}
	\hat{\lambda}_u = \mathbb{I}\left\{\sum_{i=uM_\tau N}^{(u+1)M_\tau N-1}\sum_{j=0}^{N_yN_z-1} \left|\hat H_{i,j}\right|^2>\xi_{th}\right\}, 
\end{align}
where $\xi_{th}$ is a pre-defined threshold.

The ConvSBL-GAMP is summarized in Algorithm \ref{ConvSBL-GAMP}. Note that although each device transmits the independent Gaussian pilots, the elements of the pilot matrix $\mathbf X$ are not independent due to the 2D convolution in (\ref{rpilot}), which may lead to divergence of the conventional AMP\cite{s33}. Therefore, the damping parameter $\rho$ is added in Algorithm \ref{ConvSBL-GAMP} (see \emph {lines $5,7-9$}), to guarantee the convergence of ConvSBL-GAMP. The basic computation in the ConvSBL-GAMP is the matrix multiplication leading to the complexity $\mathcal{O}(U(M_\tau N)^2N_yN_z)$ in each iteration. Compared with the complexity $\mathcal{O}((K+1)N_yN_z(UM_\tau N)^2+ 2(M_\tau N)^3UN_yN_z)$ of TDSBL-FM, the computations in ConvSBL-GAMP has been greatly reduced. Therefore, the ConvSBL-GAMP is more suitable for the large-scale systems.

\begin{algorithm}[hbt!]
	\caption{ConvSBL-GAMP Algorithm}\label{ConvSBL-GAMP}
	\begin{algorithmic}[1]
		\REQUIRE{Recieved symbols $\mathbf Y$, pilot matrix $\mathbf X$; the maximum number of iterations $T_{gamp}$, the termination threshold $\eta_{g}$, the damping parameter $\rho$, the detection threshold $\xi_{th}$, and the coupled parameter $\beta$.}
		\ENSURE The estimated channel matrix $\hat{\mathbf H}$ and the activity indicatior vector $\hat{\lambda}$.
		\STATE Initialization: $\forall i,j,o$: $\mu_{i,j}^0=0$, $\mu_{i,j}^1=10^{-8}$, $\gamma_{i,j}^0=10^{-2}$,  $\Sigma^1_{i,j} = \gamma_{i,j}^0$,  $\hat{s}_{o,j}^0=\tau_{o,j}^{p,0}=\bar{\mu}_{i,j}^{t-1}=0$; $t=0$. 
		\STATE $\backslash \backslash$ \textbf{Estimate the channel matrix}.
		\WHILE{$t < T_{gamp}$ $\&$ $\sum_{j=0}^{N_yN_z-1}\frac{\left\|\mu^t_j-\mu^{t+1}_j\right\|^2}{\left\|\mu^{t}_j\right\|^2} > \eta_g$}
		\STATE $t \gets t+1$
		\STATE $\forall o,j$: $\tau_{o,j}^{p,t}= \rho \sum_{i}\left|X_{o,i}\right|^2 \Sigma_{i,j}^t
		 + (1 - \rho) \tau_{o,j}^{p,t-1} $
		\STATE $\forall o,j$: Update $\hat{p}_{o,j}^t$ according to (\ref{pp}).
		\STATE $\forall o,j$: Update\\ $\hat{s}_{o,j}^t =\frac{\rho}{\tau_{o,j}^{p,t}} \left(\mathrm E\left[g_{o,j} | y_{o,j}, \hat{p}_{o,j}^t,\tau_{o,j}^{p,t}\right]-\hat{p}_{o,j}^t\right) + (1-\rho)\hat{s}_{o,j}^{t-1} $
		\STATE $\forall o,j$: Update\\ $\tau_{o,j}^{s,t}=\frac{\rho}{\tau_{o,j}^{p,t}}(1-\frac{1}{\tau_{o,j}^{p,t}}\mathrm{Var}\left[g_{o,j} | y_{o,j}, \hat{p}_{o,j}^t,\tau_{o,j}^{p,t}\right])+(1-\rho)\tau_{o,j}^{s,t-1}$
		\STATE $\forall i,j$: Update $\bar{\mu}_{i,j}^t=\rho \mu_{i,j}^t + (1 - \rho)\bar{\mu}_{i,j}^{t-1}$
		\STATE $\forall i,j$: Update $\tau^{r,t}_{i,j}$ and $\hat{r}_{i,j}^t$ according to (\ref{ee}) and (\ref{ee1}), where $\mu_{i,j}^t$ is replaced for $\bar{\mu}_{i,j}^t$.
		\STATE $\forall i,j$: Update $\mu_{i,j}^t$ and $\Sigma_{i,j}^t$ according to (\ref{mmm}) and (\ref{mmm2}).
		\STATE $\backslash \backslash$ \textbf{Update the hyperparameters}.
		\STATE Update the noise precision $\theta^{t+1}$ according to (\ref{theta_up}).
		\STATE $\forall i,j$: Update the local precision parameter $\alpha_{i, j}^{t+1}$ according to (\ref{aaa}) and (\ref{aaa1}).
		\STATE $\forall i,j$: Update the $\gamma_{i,j}^t$ according to (\ref{ggg}).
		\STATE $\forall u$: Update $\lambda_u^t$ according to (\ref{lab}).
		%\IF{$N$ is even}
		%\STATE $X \gets X \times X$
		%\STATE $N \gets \frac{N}{2} $  \COMMENT{This is a comment}
		%\ELSIF{$N$ is odd}
		%\STATE $y \gets y \times X$
		%\STATE $N \gets N - 1$
		%\ENDIF
		\ENDWHILE
		\STATE $\hat{\mathbf H}\gets[\mu_0^t,\dots,\mu_{N_yN_z-1}^t]$
		\STATE $\backslash \backslash $ \textbf{Detect the active device}.
		\STATE Update $\hat{\lambda}=[\lambda^t_0,\dots,\lambda^t_{U-1}]$ according to (\ref{ddd}).
		\RETURN $\hat{\mathbf H}$ and $\hat{\lambda}$.
	\end{algorithmic}
\end{algorithm}

\section{Numerical Results}
\label{4}
\begin{table}
	\scriptsize
	\caption{SIMULATION PARAMETERS}
	\begin{center}
		\begin{tabular}{|c|c|}
			\hline Parameter & Values \\
			\hline Carrier frequency $(\mathrm{GHz})$ & 2 \\
			\hline Subcarrier spacing $(\mathrm{kHz})$ & 330 \\
			\hline Size of OTFS symbol $(M, N)$ & $(256,15)$ \\
			\hline Length of CP  & 85 \\
			\hline Number of Satellite antennas & $4/16/36/64/144$ \\
			\hline Number of user terminal antennas & 1 \\
			\hline Number of LoS paths & 1 \\
			\hline Number of Non-LoS path & $3$ \\
			\hline The residual delay (\textmu s) & $[0,0.8]$\\
			\hline Doppler shift (kHz) & $[-41,41]$\\
			\hline Directional cosine along the y- and z-axis & $[-1,1]$\\
			\hline Rician factor (dB) & $5$
			\\
			\hline
		\end{tabular}
		\label{spa}
	\end{center}
\end{table}

In this section, simulations are utilized to demonstrate the performance of the proposed algorithms. We consider the senarios of the non-terrestrial networks recommended by the 3GPP\cite{3gpp} and assume that the delay of each device is precompensated in the initial downlink synchronazation so that the residual delay is in a small range. The detailed system parameters are summarized in Table \ref{spa}. 
Here, we adopt the all-ones matrix to approximate the phase compensation matrix (see (\ref{matrixform}) and (\ref{prob})) 
%The number of the adopted antennas and the pilot overhead will be 64 and 0.3, respectively, if there is no special indication. 
and set $D=a=1$, $b=r=s=10^{-4}$ and $P_{fa}=10^{-3}$ in our algorithms. Each user transmits Gaussian pliot matrix where entries in the pliot matrix are independent and obey $\mathcal{CN}(0,\frac{1}{M_\tau N})$. Moreover, the power of multi-path channel gain
is normalized as 1, i.e., $\sum_{i=0}^{p}\left|h_{u,i}\right|^2=1$. Finally, we define the signal-to-noise ratio (SNR) as $\text{ SNR }=10\log_{10}\frac{\theta}{MN}$, where $\theta$ is the precision of the Gaussian noise. 

To show the effectiveness of the proposed algorithms, we adopt SBL-GAMP\cite{s42} and GMMV-AMP\cite{s43} as benchmarks, where the SBL-GAMP was based on the conventional SBL prior only considering the single element in the channel matrix, and the GMMV-AMP explored the sparsity in the angular domain. Besides, the normalized mean square error (NMSE) and the average error probability is adopted as the metrics for the channel estimations and device activity detection, respectively, given by
\begin{align}
\text{ NMSE }=\frac{\left\| \mathbf H-\hat{\mathbf H}\right\|^2_{\mathrm{F}}}{\left\|\mathbf H\right\|^2_{\mathrm{F}}},
P_e = \frac{1}{U}\sum_{u=0}^{U-1}\left|\lambda_u-\hat \lambda_u\right|.
\end{align} 
\begin{figure}[!htb]
	\centering
	\includegraphics[width=3in]{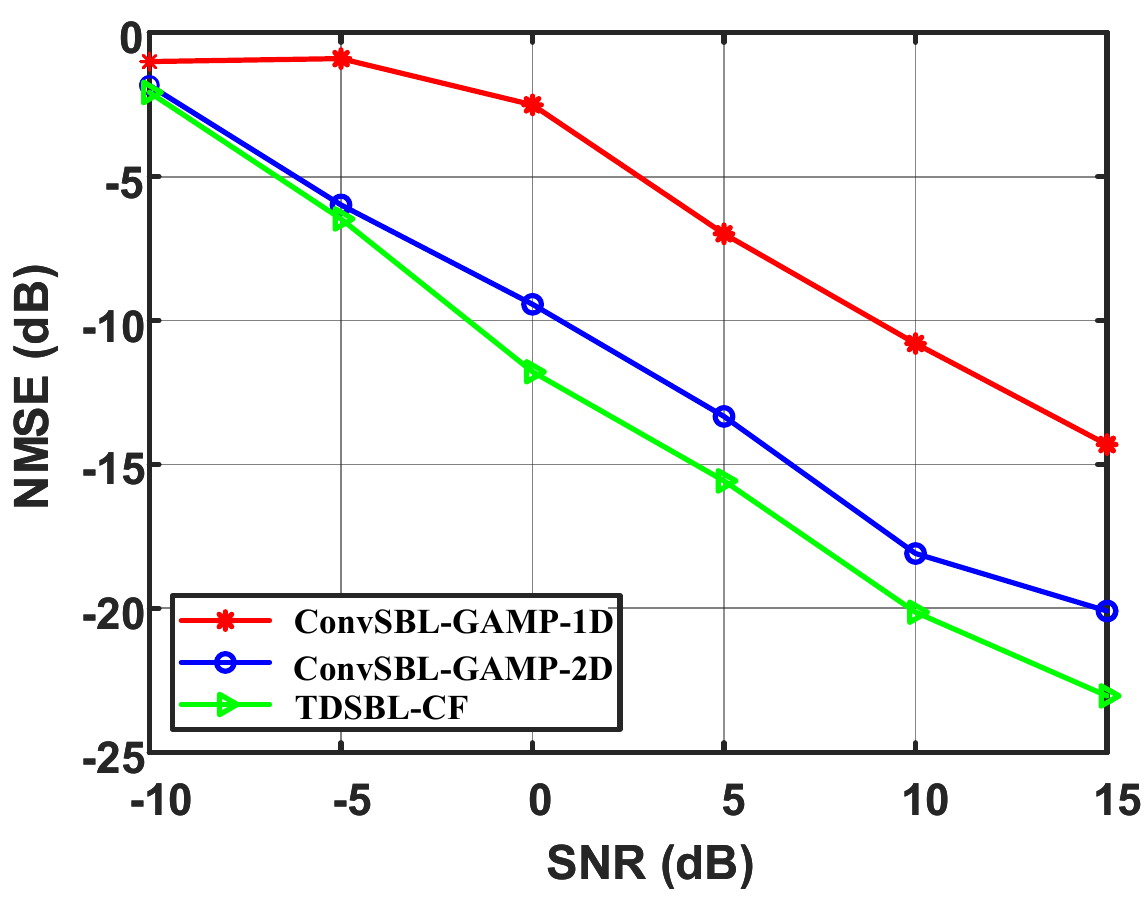}
	\caption{Performance comparison between the ConvSBL-GAMP and TDSBL-CF under different SNR values. ConvSBL-GAMP-1D (ConvSBL-GAMP-2D) represents the ConvSBL-GAMP with the one (two) dimensional kernel. $U=U_a=3$, $N_y=N_z=8$, and the  pliot overhead is 0.3.}
	\label{1d2d_com}
\end{figure} 

Firstly, as shown in Fig. \ref{1d2d_com}, we compare the performance for the channel estimation between the proposed TDSBL-CF and ConvSBL-GAMP, where the ConvSBL-GAMP-1D (ConvSBL-GAMP-2D) represents that the ConvSBL-GAMP adopts the one (two) dimensional kernel $\mathcal{B}$. For the ConvSBL-GAMP-1D, we set $\beta = 0.5$, and for the ConvSBL-GAMP-2D, we set $\beta = 0.125$, which means that the neighborhoods of the channel $h_{i,j}$ have the same influence on it. Here, we set $U=U_a=3$, $N_y=N_z=8$, and the pilot ratio is 0.3 for facilitating the computations of TDSBL-CF. %Note that although there are only three devices, the scale of this problem is much larger than that in the conventional literature, since the time-varying channel is under consideration. To see this, the dimension of the pilot matrix $X$ will be $1125\times3375$ when $\mathbf{3}$ potential devices exist, compared with the typical dimension $90 \times 2000$\cite{s33} when considering the block fading channel with the $\mathbf{2000}$ potential devices. 
In addition, we assume that the phase compensation is known here and the numerical results
are obtained by averaging over 100 channel realizations. It is observed that ConvSBL-GAMP-2D outperforms the ConvSBL-GAMP-1D for different SNR values. For example, in order to achieve the same NMSE, the SNR required by ConvSBL-GAMP-2D is about 8 dB lower than that of the ConvSBL-GAMP-1D, since the 1D kernel only accounts for the sparsity in the angular domain while the 2D kernel can capture the 2D burst block sparsity. It is indicated the effectiveness of the proposed 2D kernel.
Besides, the NMSE of the TDSBL-CF is always lower than the ConvSBL-GAMP-2D, since the TDSBL-CF computes the posterior mean of the channel more precisely and its coupled hyperparameter can be adjusted automatically in each iteration %which makes it better capture the sparsity pattern of channel. 
while ConvSBL-GAMP only adopts the single fixed coupled parameter which is less flexible. However, as we discussed earlier, with a little performance degradation, the computational complexity of ConvSBL-GAMP is far less than the TDSBL-CF, which makes it more suitable for the large-scale systems. Overall, both the TDSBL-CF and ConvSBL-GAMP are effective for estimating the channel matrix. Further, the TDSBL-CF can achieve superior performance while with more computations, and the ConvSBL-GAMP has lower computation complexity but with a little performance loss, which allows us to trade off between the complexity and performance.
In the following experiments, we show the performance comparison between the ConvSBL-GAMP and benchmarks for dealing with large dimensional problems, and omit the TDSBL-CF curve since its computation will be prohibitive in those scenarios.

\begin{figure}[!htb]
	\centering
	\includegraphics[width=3in]{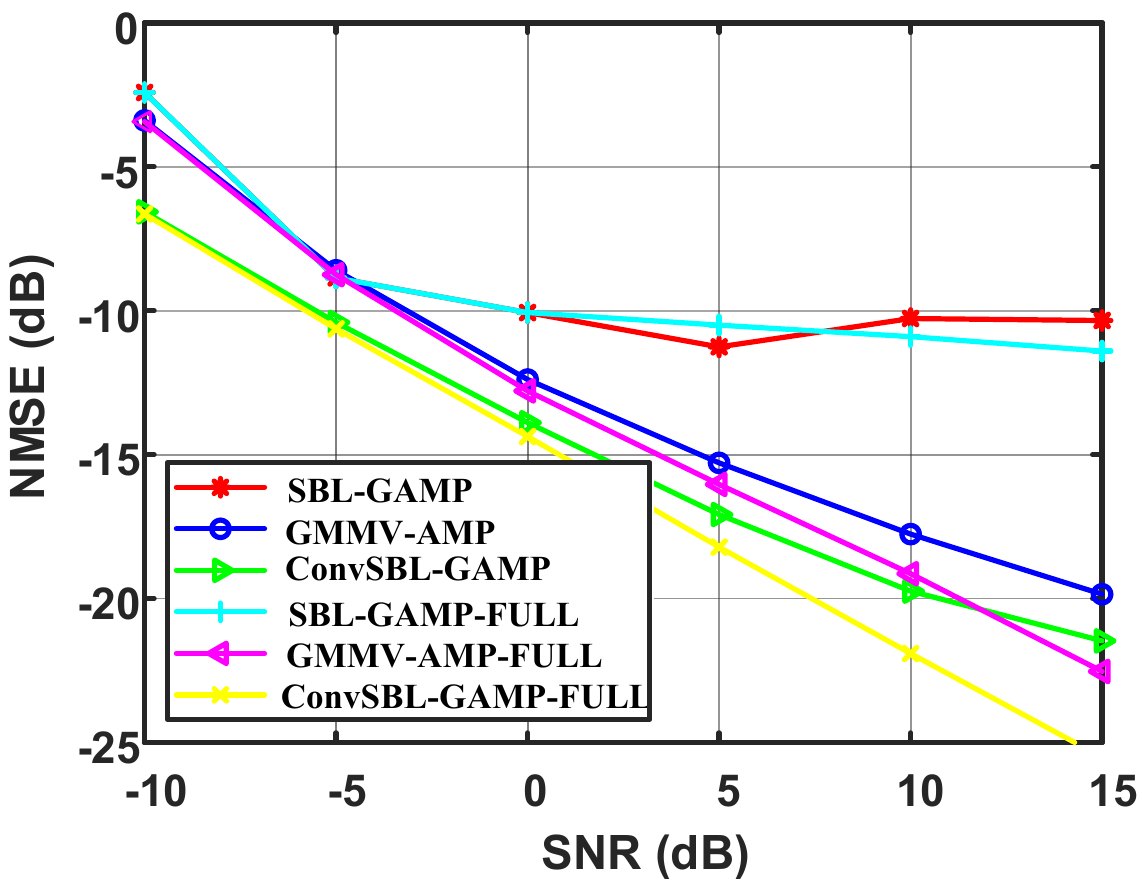}
	\caption{Performance comparison between the ConvSBL-GAMP, GMMV-AMP, and SBL-GAMP under different SNR values. $U=50$, $U_a=10$, $N_y=N_z=8$, and the pliot overhead is 0.3.}
	\label{NMSE_SNR}
\end{figure} 

Fig. \ref{NMSE_SNR} shows the NMSE comparison between the ConvSBL-GAMP, GMMV-AMP, and SBL-GAMP. The number of the potential devices and active devices are 50 and 10, respectively, i.e., $U=50$ and $U_a=10$, and the numerical results are obtained by averaging over 100 channel realizations. Here, we adopt the identity matrix to approximate the phase compensation matrix, and we also provide the performance comparison between the three algorithms when the phase compensation matrix is fully known (marked by "FULL" in Fig. 5). It is observed that ConvSBL-GAMP outperforms the other algorithms under different SNR values, even when other algorithms have the knowledge of the phase compensation matrix. For example, to achieve the NMSE of -15 dB, the SNR required by the ConvSBL-GAMP is about 3 dB lower than the GMMV-AMP and 1.5 dB lower than the GMMV-AMP-FULL, since the GMMV-AMP only captures the sparsity in the angular domain, while the ConvSBL-GAMP is able to deal with the sparsity in the angular domain and delay-Doppler domain. Notice that the NMSE of SBL-GAMP is much more than other algorithms and fluctuates around -10 dB when the SNR is larger than 0 dB, since it only accounts for the sparsity of the single element in the channel matrix, and thus fails to capture the sparsities in the angular domain and the delay-Doppler domain. Therefore, utilizing the 2D burst block sparsity is favourable for the channel estimation and our proposed algorithm can capture that feature better compared with the conventional algorithms. In addition, it is observed that the performance of the three algorithms can be improved further when the phase compensation matrix is known. However, the estimation of the phase compensation will bring extra complexity to the algorithms which may limit their applications for the large-scale systems. 
%for the ConvSBL-GAMP, we can see that there is about 0.5 dB loss when the SNR is larger than -5 dB and lower than 5 dB, which indicates that, under our system parameters, the phase compensation matrix can be approximately by the identity matrix in the low SNR regime with little performance degradation. 

\begin{figure}[!htb]
	\centering
	\includegraphics[width=3in]{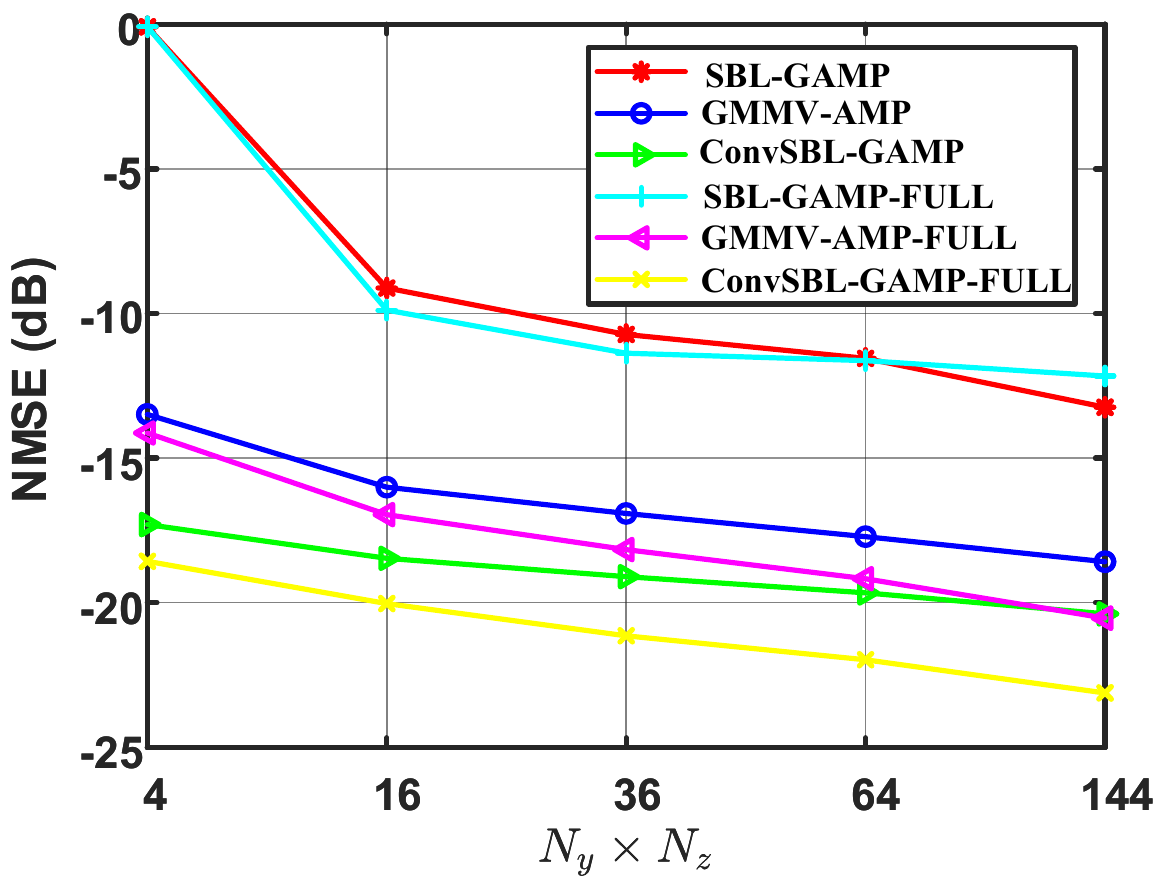}
	\caption{Performance comparison between the ConvSBL-GAMP, GMMV-AMP, and SBL-GAMP under different number of antennas. $U=50$, $U_a=10$, SNR = 10 dB, and the pliot overhead is 0.3.}
	\label{NMSE_AN}
\end{figure}

%\begin{figure}
%\centering
%\subfigure[Performance comparison under different antennas.]{
%\label{NMSE_AN}
%\begin{minipage}{4.5cm}%[b]%{0.2\textwidth}
%\includegraphics[width=\textwidth]{figure/exp/NMSE_AN50.pdf} \\

%\end{minipage}
%}
%\subfigure[Performance comparison under different pilot overhead.]{
%\label{NMSE_pilot}
%\begin{minipage}{4.5cm}%[b]%{0.2\textwidth}
%\includegraphics[width=\textwidth]{figure/exp/NMSE_pliot50.pdf} \\

%\end{minipage}
%}
%\caption{Performance comparison between the ConvSBL-GAMP, GMMV-AMP, and SBL-GAMP. $U=50$, $U_a=10$, and SNR = 10 dB.} 
%\label{11}
%\end{figure}

Next, as shown in Fig. \ref{NMSE_AN}, we provide the performance comparison for the channel estimation under the different number of antennas. Here, we set SNR = 10 dB, the pilot overhead is 0.3, and the numerical results here are obtained by averaging over 100 channel realizations. It is observed that the NMSE of the three algorithms can be decreased when more antennas are adopted, and the ConvSBL-GAMP always outperforms the other algorithms, which indicates its effectiveness. For example, the NMSE of ConvSBL-GAMP is about 3.5 dB less than that of the GMMV-AMP when the 36 antennas are adopted, and about 2.5 dB less than that of the GMMV-AMP-FULL (i.e., the phase compensation matrix is known). In addition, we can see that the NMSE of the ConvSBL-GAMP is about -17.5 dB when the 4 antennas are adopted, and is about 4.5 dB and 17 dB lower than those of the GMMV-AMP and SBL-GAMP, respectively. It is indicated that the ConvSBL-GAMP can estimate the channel matrix precisely even if fewer antennas are adopted.
\begin{figure}[!htb]
	\centering
	\includegraphics[width=3in]{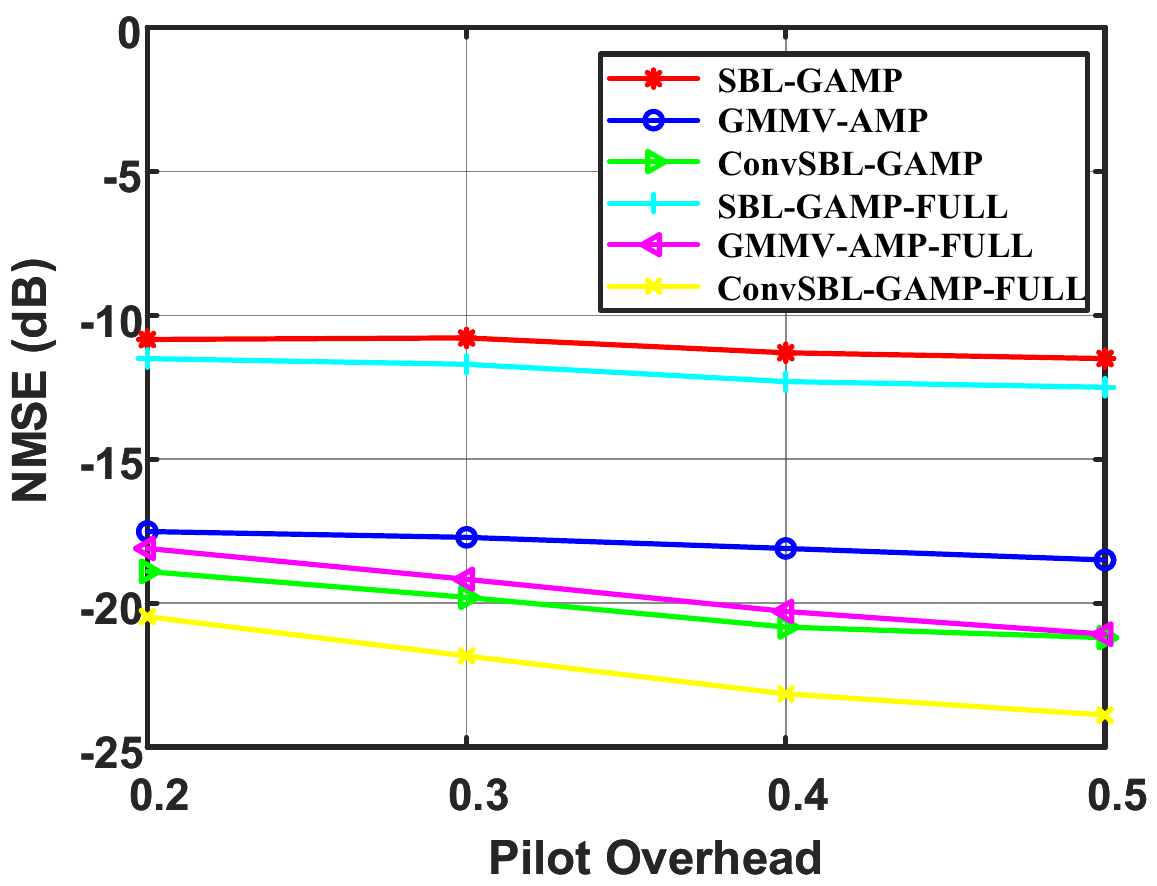}
	\caption{Performance comparison between the ConvSBL-GAMP, GMMV-AMP, and SBL-GAMP under different pilot ratios. $U=50$, $U_a=10$, SNR = 10 dB, and $N_y=N_z=8$.}
	\label{NMSE_pilot}
\end{figure} 

Then, we show the performance comparison for the channel estimation under different pilot ratios in Fig. \ref{NMSE_pilot}, where the pilot overhead is defined as the ratio between the number of pilots and total delay-Doppler domain resources, i.e., $\frac{M_\tau N_\nu}{MN}$.  Here, we set SNR = 10 dB, and 64 antennas are adopted. It is observed that the NMSE of ConvSBL-GAMP decreases with the increase of the pilot overhead and outperforms other algorithms, which indicates its superiority.  

\begin{figure}[!htb]
	\centering
	\includegraphics[width=3in]{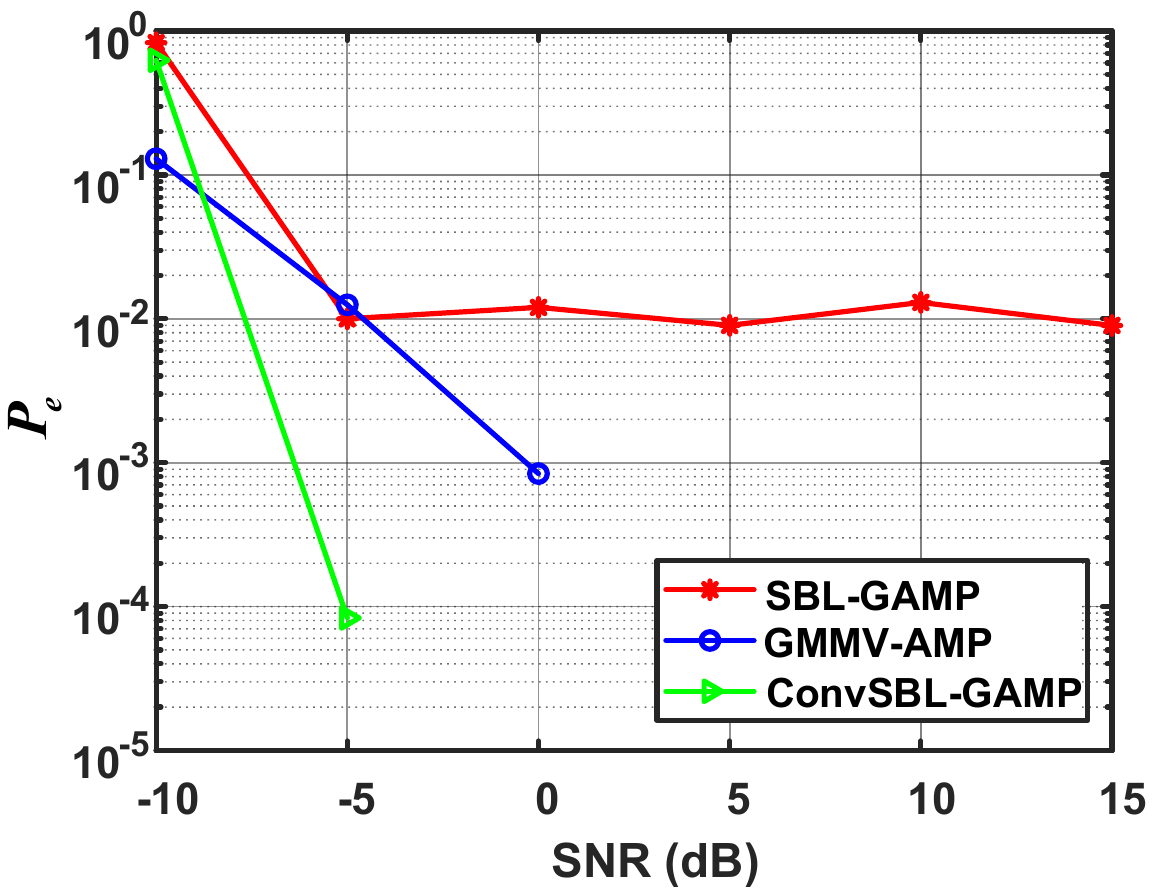}
	\caption{Performance comparison of device activity detection between the ConvSBL-GAMP, GMMV-AMP, and SBL-GAMP under different SNR values. $U=10$, $U_a=2$, the pilot overhead is 0.25, and $N_y=N_z=2$.}
	\label{NMSE_Pe}
\end{figure} 
Finally, we show the performance comparison for device activity detection between the  ConvSBL-GAMP, GMMV-AMP, and SBL-GAMP under different SNR values. The detection threshold $\xi_{th}=0.5$. There are $10$ potential devices, two of which are active, i.e., $U=10$ and $U_a = 2$. The number of antennas is set as $N_y=N_z=2$, and the pilot overhead is 0.25. The numerical results here are obtained by averaging over 1200 channel realizations. It is observed that the error probability of the ConvSBL-GAMP is very close to zero when the SNR is larger than -5 dB and is lower than the other algorithms, which indicates that the ConvSBL-GAMP can successfully detect the active devices in the low SNR regime. This may be due to the time-varying channel is considered, and the channel of each device is represented by $M_\tau N$ rows of $H$, which are much larger than that in the conventional channel matrix\cite{s33} (i.e., the channel of each device is represented by only one row of the channel matrix). Then, the difference between the energy of the channel of the active devices and inactive devices are larger than that in the conventional channel matrix, and thus the active device detection will be accurate when the channel estimation is precise enough. In addition, the error probability of the SBL-GAMP fluctuates around $10^{-2}$ when the SNR is larger than -5 dB. This may be due to the inaccurate channel estimation of the SBL-GAMP, which is consistent with that in Fig. \ref{NMSE_SNR}.
%where  SBL-GAMP adopts the conventional SBL prior which do not cpature the block sparsity; GMMV-AMP aims at recovering the channel in the angular domain and captures the burst block sparsity along the row of the channel matrix .

\section{Conclusion}
\label{con}
This work investigated the application of OTFS for the grant-free random access with massive MIMO in LEO satellite communications. The input-output relationship in the SISO-OTFS system was firstly analyzed, and then extended to the random access with massive MIMO-OTFS. Next, by exploring the two-dimensional burst block sparsity in the delay-Doppler-angle domain, the TDSBL-FM was proposed to estimate the channel and detect the active device. Then, ConvSBL-GAMP was proposed to further decrease the complexity of the TDSBL-FM. Simulation results demonstrated that the proposed scheme could acquire accurate channel state information and device activity.

\appendices
\section{Proof of Proposition 1}
\label{p1}
According to (\ref{yy}), we can get
\begin{align}
Y^{\text{TF}}&[n,m] \nonumber\\= 
&\sum_{m^{\prime}}\sum_{n^{\prime}} X^{\text{TF}}[n^{\prime},m^{\prime}] \iint h(\tau,\nu) e^{\bar{\jmath}2\pi n^{\prime}\nu T_{sym}}e^{\bar{\jmath}2\pi \nu T_{\text{cp}}} \nonumber\\ 
&A_{g_{\mathrm{rx}}, g_{\mathrm{tx}}}((n-n^{\prime})T_{sym}-\tau,(m-m^{\prime})\Delta f-\nu)\nonumber\\
&e^{\bar{\jmath}2\pi (m^{\prime}\Delta f+\nu )((n-n^{\prime})T_{sym}-\tau)}  d\tau d\nu
 \nonumber \\
%&\times   \nonumber\\ 
\overset{(a)}{=}&\frac{1}{T}\sum_{i=0}^P h_i  \sum_{m^{\prime}} X^{\text{TF}}[n,m^{\prime}] e^{-\bar{\jmath}2\pi (m^{\prime}\Delta f+\nu_i)\tau_i} e^{\bar{\jmath}2\pi n\nu_i T_{\text{sym}}} \nonumber\\	 &\int_{T_{\text{cp}- \tau_i}}^{T_{\text{sym}} - \tau_i} e^{-\bar{\jmath}2\pi((m-m^{\prime})\Delta f - \nu_i)(t^{\prime}+\tau_i)}dt^{\prime} e^{\bar{\jmath}2\pi (m-m^{\prime})\Delta f T_{\text{cp}}} 
%\nonumber\\	
%&  \nonumber \\
\end{align}
where $(a)$ is due to (\ref{cc}) and $T_{\text{ cp}} > \tau$. Thus $A_{g_{\mathrm{rx}}, g_{\mathrm{tx}}}((n-n^{\prime})T_{sym}-\tau,(m-m^{\prime})\Delta f-\nu) \neq 0$ only if $n^{\prime}=n$. Then, according to (\ref{xx}), (\ref{tap}), and (\ref{yys}), 
\begin{align}
Y^{\text{DD}}&[k,l]  \nonumber \\
=&\frac{1}{NM}\sum_{i=0}^{P} h_{i} \sum_{k^{\prime}}\sum_{l^{\prime}}X^{\text{DD}}[k^{\prime},l^{\prime}]
\frac{1}{M} \sum_{p=M_{\text{cp}}-(l_{i}+c_iM)}^{M+M_{\text{cp}}-1-(l_{i}+c_iM)} \nonumber \\ 
&e^{\bar{\jmath} 2 \pi \frac{p\left(k_i+\tilde{k}_i + b_iN\right)}{(M+M_{\text{cp}})N }} 
\sum_{m=0}^{M-1} e^{-\bar{\jmath}2\pi\frac{ m(p-l+(l_i+c_iM)-M_{\text{cp}})}{M}} 
\nonumber \\
&\sum_{m^{\prime}=0}^{M-1} e^{\bar{\jmath}2\pi\frac{ m^{\prime}(p-l^{\prime}-M_{\text{cp}})}{M} }\sum_{n=0}^{N-1} e^{-\bar{\jmath}2\pi\frac{ n(k-k^{\prime}-k_i-\tilde{k}_i-b_iN)}{N}}	\nonumber\\	
=&\sum_{i=0}^{P} h_{i} \sum_{k^{\prime}}\sum_{l^{\prime}}X^{\text{DD}}[k^{\prime},l^{\prime}] \Pi_N(k-k^{\prime}-k_i-\tilde{k}_i-b_iN) \nonumber \\  &\sum_{p=M_{\text{cp}}-(l_{i}+c_iM)}^{M+M_{\text{cp}}-1-(l_{i}+c_iM)} e^{\bar{\jmath} 2 \pi \frac{p\left(k_i+\tilde{k}_i + b_iN\right)}{(M+M_{\text{cp}})N }} \delta((p-l^{\prime}-M_{\text{cp}})_M) \nonumber \\
&\times\delta((p-l+l_i+c_iM-M_{\text{cp}})_M)  \nonumber \\
=&\frac{1}{N} \sum_{i=0}^{P} \sum_{k^{\prime}} X^{\mathrm{DD}}\left[\left\langle k-k^{\prime}\right\rangle_{N},\left(l-l_i\right)_{M}\right] 
\sum_{j=0}^{N-1} h_i e^{-\bar{\jmath}2\pi\frac{k^{\prime}j}{N}}  \nonumber\\
&e^{\bar{\jmath} 2 \pi \frac{(M_{\text{cp}}+j(M+M_{\text{cp}})-l_i)
		(k_i + \tilde{k}_i + b_iN)}{\left(M+M_{\text{cp}}\right){N}}} e^{\bar{\jmath} 2 \pi \frac{\left(k_{i}+\tilde{k}_{i} + b_iN\right)\left(l-c_iM\right)}{\left(M+M_{\mathrm{cp}}\right) N}}\nonumber\\
=&\frac{1}{N} \sum_{i=0}^{P} \sum_{k^{\prime}} X^{\mathrm{DD}}\left[\left\langle k-k^{\prime}\right\rangle_{N},\left(l-l_i\right)_{M}\right] \nonumber\\
&\sum_{j=0}^{N-1} h_{M_{\text{cp}}+j(M+M_{\text{cp}}),l_i}  e^{-\bar{\jmath} \frac{2 \pi}{N} k^{\prime} j}
e^{\bar{\jmath} 2 \pi \frac{\left(k_{i}+\tilde{k}_{i} + b_iN\right)\left(l-c_iM\right)}{\left(M+M_{\mathrm{cp}}\right) N}}
\end{align}
where $\Pi_N(x)\triangleq\frac{1}{N} \sum_{i=0}^{N-1} e^{-\bar{\jmath} 2 \pi \frac{x}{N} i}$. The proof is compeleted.
\section{Proof of Proposition 2}
\label{p2}
Computing the first derivative of the objective function in (\ref{A_f}) with respect to $\alpha_{i,j}$ and setting it equal to zero, we get
\begin{align}
\label{ppp}
	\frac{a}{\alpha_{i, j}^*}-b + \sum_{k=i-D}^{i+D}\sum_{l=j-D}^{j+D}\frac{\beta_{k,l,i,j}}{\sum_{p=k-D}^{k+D}\sum_{q=l-D}^{l+D}\beta_{k,l,p,q}\alpha_{p,q}^*} \nonumber\\-
	\sum_{k=i-D}^{i+D}\sum_{l=j-D}^{j+D} \beta_{k,l,i,j} \mathrm{E}[\left|h_{k,l}\right|^2] = 0.
\end{align}
Since $\beta_{k,l,i,j}, \beta_{k,l,p,q}, \alpha_{p,q}^*\geq0$, we have
\begin{align}
\label{ppp1}
&\sum_{k=i-D}^{i+D}\sum_{l=j-D}^{j+D}\frac{\beta_{k,l,i,j}}{\sum_{p=k-D}^{k+D}\sum_{q=l-D}^{l+D}\beta_{k,l,p,q}\alpha_{p,q}^*}\geq 0,  \\
&\sum_{k=i-D}^{i+D}\sum_{l=j-D}^{j+D}\frac{\beta_{k,l,i,j}}{\sum_{p=k-D}^{k+D}\sum_{q=l-D}^{l+D}\beta_{k,l,p,q}\alpha_{p,q}^*} \leq \frac{(2D+1)^2}{\alpha_{p,q}^*}\label{ppp2}.
\end{align}
Then, substituting (\ref{ppp}) into (\ref{ppp1}) and (\ref{ppp2}), we get the result in Proposition 2.
\bibliographystyle{IEEEtran}%
\bibliography{bibfile}

\end{document}